\documentclass[
  english,
  aps,
  prl,
  reprint,
  floatfix,
  nofootinbib,
  preprintnumbers,
  superscriptaddress,
  longbibliography
]{revtex4-2}

\usepackage{mathtools}
\usepackage{amsmath, amsfonts, amsthm, amssymb, graphicx, color, hyperref, slashed, braket}
\usepackage{subfigure}
\usepackage{pstool}
\usepackage{graphicx}
\usepackage{float}
\usepackage[utf8]{inputenc}
\usepackage[normalem]{ulem}
\usepackage[english]{babel}
\usepackage{blkarray}
\usepackage{mathrsfs}
\usepackage{diagbox}
\usepackage{multirow}
\usepackage{tikz}
\usetikzlibrary{arrows.meta,positioning}
\usepackage{physics}
\usepackage{bbold}

\usetikzlibrary{decorations.pathreplacing}

\usepackage{notes2bib}

\allowdisplaybreaks[3]

\def\d{\mathrm{d}}

\renewcommand{\Im}{\operatorname*{Im}}
\renewcommand{\Re}{\operatorname*{Re}}
\newcommand{\bbGamma}{\mathbb{\Gamma}}

\begin{document}

\title{Unitary Dual-Resonance S-matrices}

\preprint{{USTC-ICTS/PCFT-26-44}}
\preprint{{Imperial--TP--2026--cdr--4}}

\author{Claudia de Rham}
\email{c.de-rham@imperial.ac.uk}
\affiliation{Abdus Salam Centre for Theoretical Physics, Imperial College, London, SW7 2AZ, UK}
\affiliation{Perimeter Institute for Theoretical Physics, 31 Caroline St N, Waterloo, ON, N2L 2Y5, Canada}

\author{Andrew J. Tolley}
\email{a.tolley@imperial.ac.uk}
\affiliation{Abdus Salam Centre for Theoretical Physics, Imperial College, London, SW7 2AZ, UK}

\author{Zhuo-Hui Wang}
\email{wzh33@mail.ustc.edu.cn}
\affiliation{Interdisciplinary Center for Theoretical Study, University of
Science and Technology of China, Hefei, Anhui 230026, China}
\affiliation{Peng Huanwu Center for Fundamental Theory, Hefei, Anhui 230026, China}

\author{Shuang-Yong Zhou}
\email{zhoushy@ustc.edu.cn}
\affiliation{Interdisciplinary Center for Theoretical Study, University of
Science and Technology of China, Hefei, Anhui 230026, China}
\affiliation{Peng Huanwu Center for Fundamental Theory, Hefei, Anhui 230026, China}

\date{\today}

\begin{abstract}
Dual-resonance models realize crossing symmetry, Regge behavior and infinitely many resonances in a  compact analytic form but their canonical tree-level realizations fall short of genuine S-matrix unitarity. Here we explicitly construct fully unitary dual-resonance amplitudes distinct from the conventional worldsheet construction. 
The zero-width resonance towers are promoted into a physical absorptive spectrum that retains their characteristic crossing and Regge organization, while unitarizing the amplitude through a mechanism distinct from standard eikonalization. The resulting S-matrices are also local and analytic, and can contain controllable second-sheet Regge trajectories as well as substantial inelasticity. More broadly, our construction opens up a new route to charting the space of consistent strongly coupled S-matrices.
\end{abstract}

\maketitle

Dual-resonance models \cite{Veneziano:1968yb,Virasoro:1969me,Lovelace:1968kjy,Shapiro:1969km,Coon:1969oil,Bardakci:1969cs,Koba:1969rw,Kikkawa:1969qy,Martin:1969tc,Bali:1969fm,Huang:1969ef}
provided a seminal framework for understanding high-energy scattering amplitudes directly from analyticity, crossing symmetry and Regge behavior \cite{DeAlfaro:1966gxc,Igi:1967zza,Logunov:1967dy,Dolen:1967zz,Dolen:1967jr, Ademollo:1967zz,Mandelstam:1968zza,Chew:1968zz, Schmid:1968zza}. 
Their defining feature is channel duality: resonances in one channel and the crossed-channel description are two representations of the same analytic amplitude, not independent contributions to be added. This is clearly visible in the Veneziano block \cite{Veneziano:1968yb}
\begin{align}
    V_{\alpha'}(s, t)&=\frac{\Gamma(-\alpha's)\Gamma(-\alpha't)}{\Gamma(-\alpha's-\alpha't)} ,
\end{align} 
which admits a dual expansion in terms of either $s$- or $t$-channel poles
\begin{equation}
        \label{eq:dualExpan}
   V_{\alpha'}(s, t)  =\sum_{n=0}^{\infty} \frac{(\alpha' s+1)_n}{n!(n-\alpha' t)}
     =\sum_{n=0}^{\infty} \frac{(\alpha' t+1)_n}{n!(n-\alpha' s)}, 
\end{equation} 
with $s,t$ and $u=-(s+t)$ being the standard Mandelstam variables and $(a)_n\equiv \Gamma(a+n)/\Gamma(a)$ being the Pochhammer symbol. This duality ultimately led to string theory. Yet, viewed as physical S-matrices, without string loop corrections, these dual models have a basic shortcoming: they are {\it not unitary}. Being tree-level meromorphic amplitudes, their absorptive parts are supported only on isolated poles, and hence they do not satisfy {\bf full partial-wave unitarity\footnote{Throughout this paper, {\it full unitarity} refers to exact partial-wave unitarity of the four-point amplitude.}}. In the context of hadron scattering, the Veneziano amplitude only provides a qualitative description of the strong-interaction data.

This leaves an important gap in the modern S-matrix bootstrap \cite{Paulos:2016fap,Paulos:2017fhb,deRham:2017avq,Tolley:2020gtv,Caron-Huot:2020cmc,Arkani-Hamed:2020blm,Bellazzini:2020cot,Hebbar:2020ukp,Sinha:2020win,Remmen:2020vts,Zhang:2020jyn,Li:2021lpe,Caron-Huot:2021rmr,Chiang:2021ziz,Guerrieri:2021ivu,Alberte:2021dnj,Haring:2022cyf,Huang:2022mdb, Henriksson:2022oeu,Antunes:2023irg,Haring:2023zwu,Cheung:2023uwn,Geiser:2023qqq,  Karateev:2019ymz,He:2023lyy,Albert:2023seb,Bhat:2024agd,Caron-Huot:2024lbf,Cheung:2024uhn,Eckner:2024ggx,Guerrieri:2024ckc,Guerrieri:2024jkn,Wan:2024eto,Saha:2024qpt,Bellazzini:2025bay,Berman:2025owb,Cheung:2025krg,Correia:2025ozf,Correia:2025uvc,He:2025gws,deRham:2025vaq,Alday:2025pmg,Elvang:2026pmc,Chang:2026ztn,Calisto:2026stl,Cheung:2026lpv,EliasMiro:2026utl,Wan:2026pjq,Gumus:2026mhb,Boisvert:2026sfh,Berman:2026ezk}, which routinely imposes fundamental principles such as unitarity, analyticity and locality, but explicit amplitudes realizing all of them while retaining the dual-resonance structure are rare. Extending the dual-resonance framework beyond tree-level string amplitudes would provide a new analytic handle on consistent dual S-matrices and useful new tools for strongly coupled theories.

In this Letter, we construct a novel class of fully unitary dual-resonance models by smearing (zero-width) dual-amplitudes over a continuous Regge slope, converting the infinite tower of poles into physical branch cuts while preserving the underlying crossing and Regge behavior. For definiteness, we focus on identical scalar amplitudes that are triple crossing-symmetric, with massless exchange poles in the $s$, $t$, and $u$ channels. The conditions required on the smearing function are derived by analyzing the relevant asymptotic regimes of the proposed dual amplitude. We verify both analytically and numerically that the resulting amplitudes define local, analytic and unitary S-matrices, with controllable second-sheet resonance trajectories and, in general, substantial inelasticity. One can further vary the smearing weights to obtain a wide family of fully unitary dual-amplitudes, ranging from resonance-dominated models with Regge-organized second-sheet poles to purely nonresonant absorptive backgrounds. This provides a concrete starting point for a broader S-matrix bootstrap program to construct consistent dual-amplitudes with strong inelasticity.

\vspace{0.1cm}

\noindent{\bf Unitary dual models:} Full unitarity requires the presence of branch cuts along the real axes of the Mandelstam variables. A natural way to generate such cuts from poles is to smear them by integration\footnote{For example, the Cauchy transform of a continuous line segment of poles gives $F(z)=\int_0^1 \frac{dx}{z-x}=\log z-\log(z-1)$.}
much as loop amplitudes arise from tree-level propagators integrated over continuous momenta. Our strategy to construct unitary dual amplitudes is to start with 
Veneziano amplitudes and convolve them with an appropriate weight function $\phi(y)$ over a continuous range of Regge slopes $y$
\begin{align}
\label{eq:Mansatz}
    \mathcal M(s,t)
    =
    \!\int_{\alpha'}^{\infty}&\!\!\d y\, \phi(y)
    W_y(s, t,u),\\
    W_y(s, t,u)&\equiv V_y(s, t)+V_y(u, t)+V_y(s, u) .\nonumber
\end{align}
The lower cutoff is $\alpha'$ (rather than $0$) so as to effectively set a maximum string tension in the spectral distribution.
The Feynman $i\epsilon$ prescription is implied in the physical region. For example, in the $s$-channel ($-s<t<0$), $s$ should be understood as $s+i\epsilon$. With this prescription, the Veneziano poles are shifted away from the real $y$-integration contour, so that the integral is well defined. As will become clearer later, the amplitude in Eq.~\eqref{eq:Mansatz} is not valid for all complex $s$ and $t$, and an analytic continuation is necessary in some regions. 

Much of the analysis is valid in arbitrary $D$ dimensions, including $D=4$, which is relevant for hadron scattering and is explored in depth in \cite{upcomingPaper} alongside unitary dual models in other dimensions, with internal symmetries, or with higher spins. In this Letter, however, we focus on $D=10$ for the sake of concreteness.   A simple family of weight functions that lead to fully unitary dual resonances is given by
\begin{equation}
\label{eq:fidModel}
    \phi(y)=C \alpha'^2 \left(1-\frac{\alpha'}{y}\right)^\beta \left(\frac{y}{\alpha'}\right)^{-p+2}\frac{\bbGamma^2}{(y-y_0)^2+\bbGamma^2}\,,
\end{equation}
where $1<p\leq 2$, $\beta>0$, $y_0>\alpha'$, $C$ is a dimensionless (positive) constant and $\bbGamma$ is a positive constant with the same dimension as $\alpha'$. 
For this simple family, we find by analyzing its partial-wave expansion that the $\ell$-wave resonance poles on the second Riemann sheet are located at
\begin{equation}
    s^{(\ell)}_{n} = \frac{n}{y_0+i\bbGamma} = \frac{n y_0}{y_0^2+\bbGamma^2} - i\frac{n \bbGamma}{y_0^2+\bbGamma^2} ,~~   \ell\leq 2\lfloor \frac{n}{2}\rfloor 
\end{equation}
with the decay width\footnote{Note that for large $n$, this implies that the widths grow as $\sqrt{n}$. Yet such a growth is consistent with the Froissart bound as checked explicitly.} given by $\Gamma_n^{\rm decay}\equiv -2 \operatorname{Im} (s_n^{(\ell)})^{1/2}=\sqrt{2n}\big[((y_0^2+\bbGamma^2)^{1/2}-y_0)/(y_0^2+\bbGamma^2)\big]^{1/2}$. 
The selection rule $\ell\leq 2\lfloor n/2\rfloor$ is inherited from the seed Veneziano amplitude, since the $y$ smearing leaves the angular dependence of each level-$n$ residue unchanged. Some spins allowed by this bound may nevertheless be absent at particular levels, exactly as in the seed amplitude.

To see this, note that these second-Riemann-sheet poles are essentially pinch singularities of the $y$-integration contour in Eq.~\eqref{eq:Mansatz}. Recall that the $s$-channel poles of $W_y(s,t,u)$ are located at $y=n/s$, which can be seen by examining the partial-wave amplitudes. For our massless scattering, we define the second-Riemann-sheet partial-wave amplitude by analytically continuing through the first cut connecting to the origin of the $s$ plane, $0<s<1/\alpha'$. As $s$ is continued to the second Riemann sheet, the $s$-channel poles $y=n/s$ move into the upper half $y$-plane. When one of these moving poles hits the upper-half-plane pole of $\phi(y)$: ${n}/{s}=y_0+i\bbGamma$, the $y$ contour is pinched, producing a resonance pole in the dual model on the second sheet.

It is worth noting that the uniform spacing of the resonance poles in the $s$-plane, together with the fixed decay-width-to-mass ratio, is reminiscent of the behavior of open-string resonances once loop corrections are included (see for example \cite{Eberhardt:2022zay, Banerjee:2024ibt}), even though the present unitarization mechanism differs in the detailed structure of its long-range behavior. Also, unlike standard eikonalization, which is primarily a high-energy, small-angle resummation and becomes compulsory for gravitational spin-2 exchange, our mechanism unitarizes scalar dual-resonance amplitudes by replacing zero-width pole towers with physical absorptive cuts.

The requirement of $\beta>0$ is mandated by the convergence of the smearing integration when the $y$ poles of the Veneziano kernels coincide with the lower limit, $\alpha'$, of the integration, while $p>1$ is required so that the smearing integration converges for large $y$. The condition $p\leq 2$ is determined by imposing partial-wave unitarity near threshold. Specifically, unitarity constrains the asymptotic behavior of $\phi(y)$ because the threshold behavior of $\Im a_\ell(s)$ is controlled by the large $y$ part of the smearing integral, which can be seen by using the pole representation of the Veneziano kernel \eqref{eq:dualExpan}; see Appendix B. Apart from the threshold,  the large-energy and large-spin limit as well as the finite-energy and large-spin limit are also sensitive to the large $y$ part of the smearing integral, and thus partial-wave unitarity in these limits also imposes constraints on the asymptotic behavior of $\phi(y)$. However, they are weaker than the threshold constraint; see Appendices C and D.

\vspace{0.1cm}

\noindent{\bf Preserving Full Unitarity:} While unitarity being satisfied in various asymptotic regions is encouraging, it is central to verify partial-wave unitarity at finite energies and spins.

\begin{figure}[t]
    \centering
    \includegraphics[width=1\linewidth]{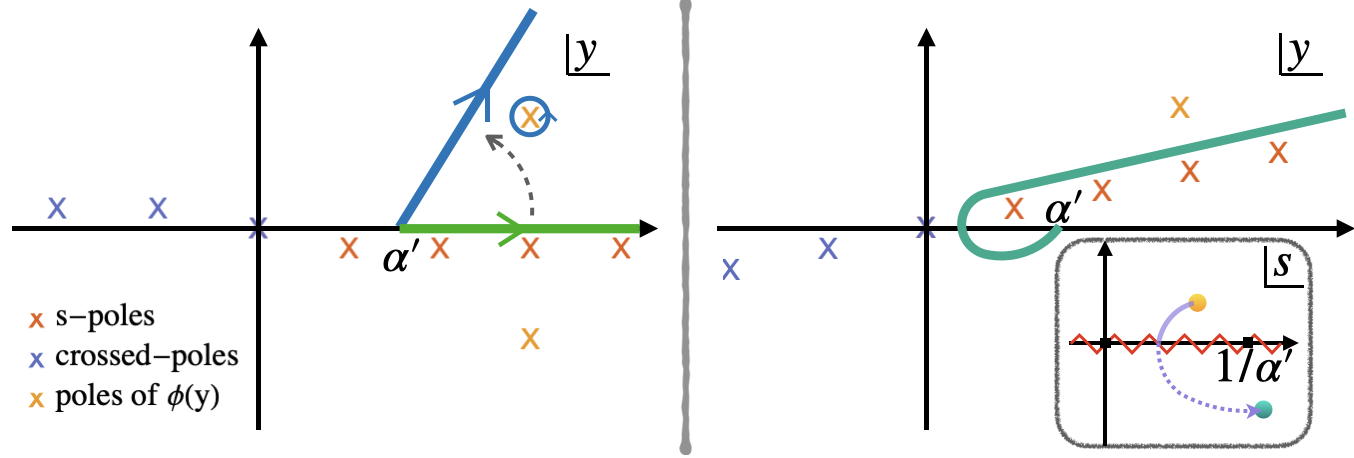}
    \caption{({\it Left}) Deformation of the $y$ integration contour of amplitude \eqref{eq:Mansatz}, making numerical integration practical. 
    ({\it Right}) Continuation through the cut $0<s<1/\alpha'$, causing $\phi(y)$'s pole to pinch with a Veneziano pole to generate second-sheet resonance poles.
    }
    \label{fig:iu_deform_y}
\end{figure}

To evaluate the amplitude \eqref{eq:Mansatz}, the Feynman $i\epsilon$ prescription implies that we must integrate above the real axis in the complex $y$ plane, avoiding the poles at $y={n}/{s},~n\in \mathbb{Z}_{>0}$. A convenient choice that ensures fast convergence is to rotate the contour around $y=\alpha'$ anti-clockwise by an angle of $\varphi_y=\pi/3$, which additionally picks up a pole contribution originating from the weight function $\phi(y)$.

The choice is suited for our use of the double-exponential method for the numerical integration, which handles endpoint singularities, although its precision is sensitive to the distance between the poles and the integration contour. More explicitly, we use the double-exponential map $y=\alpha' +2\alpha' e^{i\varphi_y}\exp\left({\pi\over 2}\sinh w_y\right)$, and sample uniformly along the real $w_y$ axis with spacing $h_y=2^{-9}$. The upper end of the $w_y$ range is truncated once its contribution to the sum falls below $O(10^{-150})$. For the later partial-wave projection, the integrand contains branch cuts in the complex $z$-plane. The double-exponential method is therefore again useful, and we set $z=\tanh\left({\pi\over 2}\sinh w_z\right)$, with precision settings similar to those used above.

The $s$-channel partial-wave amplitude can be obtained by using
\begin{equation}
   \! \!a_\ell(s)=\frac{\mathcal{N}_D}{2}\!\! \int_{-1}^1 \!\!\! \d z\frac{\left(1-z^2\right)^{\frac{D-4}{2}}}{s^{\frac{4-D}{2}}}\! P_\ell^{(D)}\!(z) {\cal M}\Big(\!s, \frac{s}{2}(z\!-\!1)\!\Big)\,,
\end{equation}
where the coefficients are normalised so that $|1+ia_\ell|\le 1$,
with $z\equiv \cos\theta$, 
$
    P^{(D)}_\ell(z) =
    {}_2F_1\!\left(
        -\ell,\ell+D-3;\frac{D-2}{2};\frac{1-z}{2}
    \right),
$ and $\mathcal{N}_D={(16 \pi)^{\frac{2-D}{2}}}/{\Gamma\left(\frac{D-2}{2}\right)}$.
Partial-wave unitarity must be satisfied in the physical region (for even $\ell$)
\begin{equation}
    0\leq Q_\ell(s)
    \equiv
    \frac{|a_\ell(s)|^2}
    {2\,\operatorname{Im}a_\ell(s)} \leq 1,~~~~ s> 0 .
\end{equation}
At threshold, phase-space suppression gives $a_\ell=0$, so the unitarity inequality is trivial. Since the Veneziano kernel has positive partial-wave residues for $D\leq 10$ and we choose a positive weight function, $\operatorname{Im}a_\ell(s)$ is obviously positive, and we only need to verify that $Q_\ell(s)\leq1$.

For the $D=10$ explicit model \eqref{eq:fidModel}, we choose  $\beta=3$, $p=2$, $y_0=2\alpha'$, and $\bbGamma=3\alpha'/5$ to present numerical results. $C=1.84\times10^7$ is chosen to approximately saturate partial-wave unitarity for the most constraining partial wave, which occurs near $s\approx1.8/\alpha'$. 

\begin{figure}[t]
    \centering
    \includegraphics[width=1\linewidth]{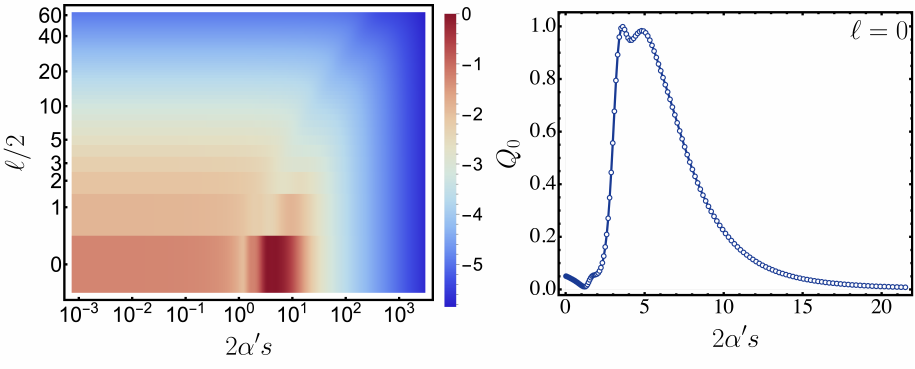}
    \caption{
   The left panel displays $\log_{10}Q_\ell(s)$ for finite $s$ and $\ell$ in the explicit model \eqref{eq:fidModel}. Partial-wave unitarity is satisfied in all regions. The right panel shows the $\ell=0$ partial wave.  We have chosen $C=1.84\times10^7$, $\beta=3$, $p=2$, $y_0=2\alpha'$, and $\bbGamma=3\alpha'/5$.
    }
    \label{fig:scalarQ}
\end{figure}

In Figure~\ref{fig:scalarQ}, we see that partial-wave unitarity is satisfied in all sampled $s$ and $\ell$.
For the explicit model \eqref{eq:fidModel}, low spin dominance is manifest, and the $\ell=0$ partial wave saturates the unitarity bound at $s\sim O(1/\alpha')$. A key feature is that partial waves are well within the unitarity bounds for large and finite $s$ and $\ell$. This, together with the asymptotic regimes we have probed, indicates that partial-wave unitarity is satisfied for all $s$ and $\ell$.

From the Regge-theory point of view, a Veneziano amplitude $W_y(s,t,u)$ with fixed $y$ gives rise to $t$-channel partial-wave amplitudes with only Regge poles in the complex $\ell$ plane, which violates unitarity \cite{Amati:1962nv, Mandelstam:1963cw}. The $y$ integration smearing upgrades the Regge poles to Regge cuts and thus resolves the unitarity violation. For example, for $\phi(y)\sim (y-\alpha')^\beta$ with integer $\beta$, the leading/rightmost $\ell$-plane pole
$a^{y}_\ell(t)\sim {1}/({\ell-yt})$ is upgraded to a Regge cut of the form
\begin{align}
a_\ell(t)&\sim \int_{\alpha^{\prime}}^{\infty} \!\!\!\d y \phi(y) a^y_\ell(t) 
\sim 
\left(\ell-\alpha^{\prime} t\right)^\beta \log \left(\ell-\alpha^{\prime} t\right) .
\end{align}
Note that, in a bootstrap setup, one may change the locations and types of the Regge cuts by adjusting the lower integration limit and $\phi(y)$'s behavior near it.

\vspace{0.1cm}

\noindent{\bf Locality:} Our dual amplitude is local in the sense that it is polynomially bounded at large $s$ for fixed $t<0$. This would be immediate if $y_{\rm max}$ were finite, since the Veneziano kernel is itself local. The analysis in Appendix C indicates that, in the Regge limit $|s|\gg -t$, one has, for example, $|V_y(s,t)|\sim y^{-1/2}\exp\left[t\log(\alpha'|s|)y\right]$ when $y$ is large. Since $t<0$, the large-$y$ region of the integral is exponentially suppressed and therefore subdominant. It follows that our dual amplitude is local. 

It actually preserves the Regge behavior of the Veneziano kernel, up to logarithmic corrections. To see this, note that for fixed $t<0$ and $|s|\gg -t$,  the $y$ integral is dominated by the $(-yt)^2/(y|s|)\ll 1$ region, and we have 
\begin{equation}
    {\cal V}(s,t) \equiv \int_{\alpha'}^{\infty}\!\!\!\d y\, \phi(y)V_y(s, t) \sim\! \int_{\alpha'}^{\infty}\!\!\!\d y\, \Gamma(-y t) \phi (y) (-y s)^{yt}\,.\nonumber
\end{equation}
For a sufficiently large $s$, we have $-1/(\alpha't)\ll\log(\alpha' |s|)$, in which case the endpoint region $y\sim \alpha'$ controls the integral. Substituting $\phi(y\to \alpha'^+)\sim (y-\alpha')^\beta$, we get
\begin{align}
    {\cal V}(s,t)\sim \Gamma(-\alpha' t )(-\alpha's)^{\alpha' t}
    \left[-\alpha't \log(-\alpha's)\right]^{-(\beta+1)} .
\end{align}
Here $\log(-\alpha's)$ is defined with a cut along $s\geq0$ and is real for $s<0$. Thus, the ${\cal V}(s,t)$ block is polynomially bounded in the complex $s$ plane for fixed $t<0$, with only logarithmic corrections. The logarithmic corrections are exactly due to a Regge pole being smeared into a $(\beta+2)$-Reggeon branch cut (if $\beta$ is an integer). The same argument applies to $V_y(u,t)$ after replacing $s$ by $u$, because $|u|\sim |s|$ in the fixed-$t$ Regge limit. For $V_y(s,u)$, a minor refinement is required, but it leads to a slightly softer Regge behavior.  

\vspace{0.1cm}

\noindent{\bf Maximal analyticity:} We now show that our dual amplitude is maximally analytic in the Mandelstam sense. Since the three blocks of the amplitude are simple crossings of each other, it suffices to prove maximal analyticity for one block, say, $\mathcal{V}(s,t)$. That is, we show that $\mathcal{V}(s,t)$ is analytic in $(\mathbb C\setminus[0,\infty))_s\times(\mathbb C\setminus[0,\infty))_t$. We note that this cannot be directly inferred from Eq.~\eqref{eq:Mansatz}, since the $y$ integral in that representation does not converge for all $s$ and $t$ in this domain.

We start with the worldsheet/Euler-Beta representation of the Veneziano block, which converges when $\Re s,\,\Re t<0$,
\begin{equation} \label{eq:continue_V_start}
    \mathcal{V}(s,t)=\int_0^1\frac{\d x}{x(1-x)}F_\phi(A_{s,t}(x))\,,
\end{equation}
where we have defined $F_\phi(A)\equiv \int_{\alpha'}^\infty \d y\,\phi(y)e^{-Ay}$ and $A_{s,t}(x)\equiv s\log x+t\log(1-x)$. For $\Re s,\,\Re t<0$, $\Re A_{s,t}(x)>0$ on $0<x<1$, allowing the exchange of $x$ and $y$ integrations; Eq.~\eqref{eq:continue_V_start} thus gives the analytic germ of the original representation. First, note that with our choice of the smearing function, $F_\phi(A)$ can be continued to the entire complex $A$-plane except for a finite branch point at $A=0$. To see this, for any $A\neq0$, one may deform the $y$-contour, avoiding the singularities of $\phi$, so that it approaches infinity along a direction where $e^{-Ay}$ decays. On the other hand, when $A$ is analytically continued around the origin, due to the $y$ singularities in $\phi(y)$, the continued contour does not return to the original one, implying that $F_\phi(A)$ has a branch point at $A=0$\,\footnote{For example, for $\phi(y)=y^{-p}(1-\alpha'/y)^\beta$, $F_\phi(A)\propto e^{-\alpha' A}U(\beta+1,2-p,\alpha' A)$, where $U$ is the Tricomi confluent hypergeometric function, which has precisely this singularity structure.}. 
Equivalently, this branch point can be read off directly from the large-$y$ behavior: if $\phi(y)\sim\sum_k c_k y^{-p-k}$ as $y\to\infty$, then the infinite part of the Laplace transform contains terms $\int_R^\infty dy\, y^{-p-k}e^{-Ay}=A^{p+k-1}\Gamma(1-p-k,AR),$ which generate $A^{p+k-1}$ or, for integer exponents,
$A^{p+k-1}\log A$ near $A=0$.
This branch point is actually a soft one with a finite limiting value, because the integration over $y$ converges when $A=0$. Another fact that we will use is that for $\Re A>0$ on the first Riemann sheet and $A\to\infty$, the $y$ integral defining $F_\phi(A)$ is controlled by the threshold behavior near $y=\alpha'$, which implies
\begin{equation}
\label{eq:FphiAapprox}
    F_\phi(A)\sim A^{-\beta-1}e^{-\alpha' A}.
\end{equation}
Let us now analytically continue $\mathcal{V}(s,t)$ from $\Re s,\Re t<0$ to general complex $s$ and $t$, by deforming the $x$ contour. Let us first deform the portions of the contour near the end points $x=0,1$. Near $x=0$, it is simplest to make a change of variable $x\to L=\log x$ and deform the $L$ contour instead. Obviously, we must choose the $L$ contour, parametrized as $L\simeq  - R e^{i \vartheta}$, to end at $x=0$, which requires $\Re e^{i \vartheta}>0$ when $R\to+\infty$. By Eq.~\eqref{eq:FphiAapprox} and $\d x /x =\d L$, a valid deformation also requires $\Re(-s e^{i \vartheta})>0$, in order for the deformed integral to converge near $x=0$. Thus, such an $L$-contour deformation near $x=0$ is valid precisely when $s\notin[0,\infty)$. Although this portion of the contour is simply a straight ray in the $L$-plane, from the $x$-plane perspective it spirals around $x=0$ whenever $e^{i\vartheta}$ has a nonzero imaginary part. The contour near  $x=1$ can be treated analogously, and the corresponding deformation is unobstructed precisely when $t\notin[0,\infty)$. Therefore, the endpoint singularities single out exactly the $s$- and $t$-channel cuts.

Next, we must show that the remaining portion of the $x$ contour is free of pinch singularities; that is, the branch point of $F_\phi(A)$ at $A=0$ does not pinch the $x$ contour as $s$ and $t$ are continued away from the $s$- and $t$-channel cuts. For this purpose, it suffices to show that the Landau equations
\begin{equation}
    A_{s,t}(x_\ast)=0
    \qquad\text{and}\qquad
    \partial_x A_{s,t}(x_\ast)=0
\end{equation}
have no solutions on the relevant contour. To see this, first note that for $s+t=0$ and $s\neq0$, $\partial_x A_{s,t}(x)=s/[x(1-x)]$ has no finite zero and hence produces no internal pinch. For $s+t\neq0$, the second equation above gives $x_\ast=s/(s+t)$, and substituting this into the first equation gives
$
    x_*\log x_* + (1-x_*)\log(1-x_*)=0.
$
Since $F_\phi(A)$ has a finite limiting value at $A=0$, a simple preimage $A_{s,t}(x)=0$ can be avoided by an arbitrarily small detour without a finite deformation of the contour. The derivative $\partial_x A_{s,t}$ is log-branch independent, so, for each fixed $s,t\in\mathbb C\setminus[0,\infty)$ with $s+t\neq0$, any finite solution of $\partial_x A_{s,t}=0$ projects to $x_\ast=s/(s+t)\notin\{0,1\}$; the endpoint windings can then be confined to sufficiently small neighborhoods of $x=0,1$ that avoid this point. Hence a possible pinch can only occur on the middle contour, where $x_\ast\log x_\ast+(1-x_\ast)\log(1-x_\ast)=0$, which is impossible because the left-hand side is strictly negative for $0<x_\ast<1$ \footnote{Only Landau solutions lying on the contour can pinch it. Since the endpoint deformations are confined near $x=0,1$, solutions with $x_\ast\notin(0,1)$ remain separated from the contour; the branch of $F_\phi$ is followed continuously.}.
Since the continuation is unobstructed throughout the simply connected domain $(\mathbb C\setminus[0,\infty))_s\times(\mathbb C\setminus[0,\infty))_t$, the resulting analytic continuation is unique and independent of the continuation path.

\vspace{0.1cm}

\begin{figure}[htbp]
    \centering
    \includegraphics[width=1\linewidth]{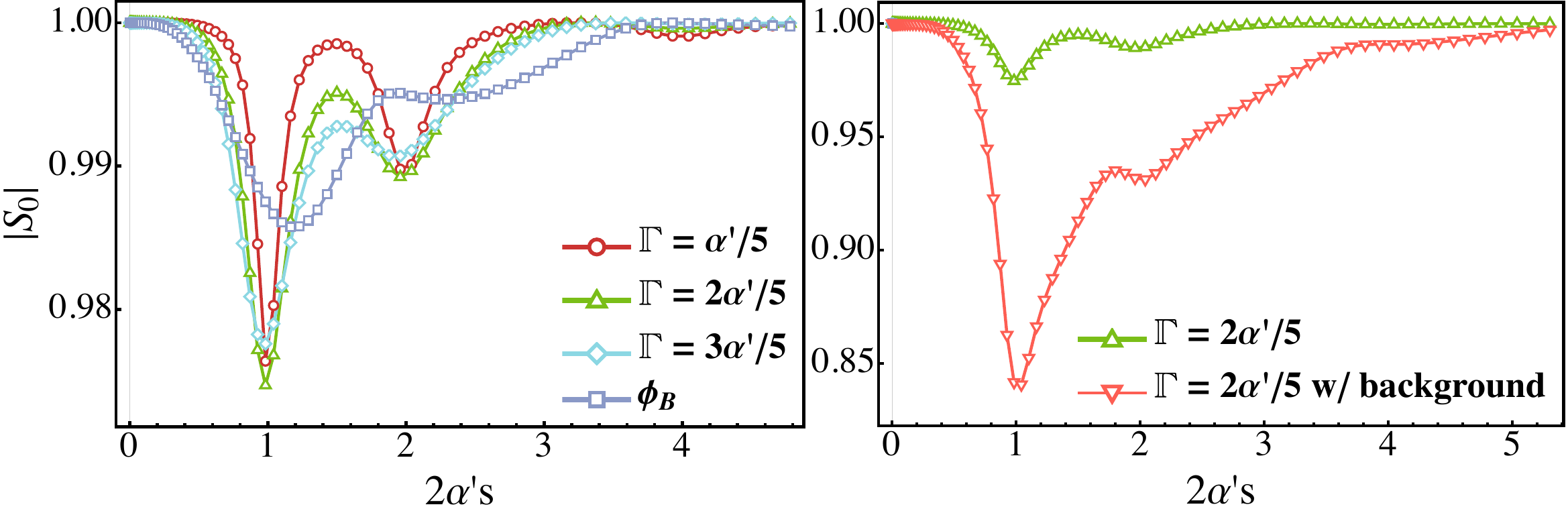}
    \caption{
     (\textit{Left}) $S_0=1+ia_0$ for several values of $\bbGamma$ with $\beta=3$, $p=2$ (see Eq.~\eqref{eq:phiBBackground} for absorptive background $\phi_B$).
    (\textit{Right}) Combining the dual-resonance amplitude \eqref{eq:Mansatz} with the Virasoro-Shapiro-like absorptive background significantly enhances inelasticity/coupling strength.}
    \label{fig:iu_addBG}
\end{figure}

\noindent{\bf Dual models without resonances:} It is interesting to note that an alternative weight function
\begin{equation}
\phi_B(y)=B \alpha'^2 (1-\alpha'/y)^\beta(y/\alpha')^{-p} \,,
\label{eq:phiBBackground}
\end{equation}
without complex $y$ poles still defines a valid S-matrix, provided that $B$ lies below a suitable upper bound. The ${\cal M}$ amplitude constructed with $\phi_B(y)$ still has unitary asymptotic limits. Moreover, it can be shown that the amplitude remains unitary at finite $s$ and $\ell$, with the largest value of $Q_\ell$ occurring around $s\sim {\cal O}(1/\alpha')$ for the $\ell=0$ partial wave, similar to that in Figure \ref{fig:scalarQ}. However, $\phi_B(y)$ is holomorphic in the upper half plane, so there are no second-Riemann-sheet resonance poles at all in this model. To see this, recall that the would-be resonance poles arise as pinch singularities in the $y$ integration, but there is no pole in the smearing function $\phi_B(y)$, with which the poles of $W_y(s,t,u)$ can form a pinch singularity. Moreover, for complex $s$ at fixed $-1\leq z\leq 1$, the deformed $y$ integration converges. Nevertheless, as we can see in Figure \ref{fig:iu_addBG}, the partial-wave amplitudes still feature resonance-like peaks, and we shall refer to $\phi_B(y)$ as an absorptive background. In this case, if we perform a phase shift analysis, the observed phase shift jumps do not correspond to the resonance poles in the second-Riemann-sheet sense.

This situation is reminiscent of ordinary loop amplitudes in weakly coupled UV completions: they generate the required branch cuts but do not by themselves generate second-Riemann-sheet resonance poles. A simple example is a two-scalar theory ($\chi_l,\,\chi_h$) with a mass hierarchy and no cubic interactions, where the leading light amplitude $\chi_l\chi_l\to\chi_l\chi_l$ is produced by a heavy-particle loop. At fixed perturbative order, the amplitude contains threshold cuts but no resonance poles. Non-perturbatively, however, this conclusion may change. This can be seen by resumming the $s$-channel bubble-chain diagrams in the $\chi_l\chi_l \to \chi_l\chi_l$ scattering, which produces a denominator in the $\ell=0$ partial wave, and thus a second-sheet pole. If the diagonal self-interactions of the two scalars are somehow tuned to be loop-suppressed, then there is no pre-existing bound state in the heavy channel. In that case, a resonance pole in this minimal off-diagonal model requires the transition coupling to be of order one in partial-wave units, {\it i.e.}, a genuinely strong-coupling regime.

\vspace{0.1cm}

\noindent{\bf Enhanced inelasticity:}
Combining a resonant smearing function with such an absorptive background yields a richer class of unitary S-matrices. It can increase both the allowed interaction strength and the inelasticity, while leaving the resonance pattern essentially intact. In the model \eqref{eq:fidModel}, the allowed scattering strength is moderate, because increasing the scattering strength readily drives the spin-0 partial wave to saturate partial-wave unitarity near $s=3/(2\alpha')$. However, by adding a Virasoro-Shapiro-like absorptive background term,
\begin{equation}
\label{eq:VSBackground}
\hspace{-0.5cm}\int_{\alpha'}^\infty \hspace{-0.2cm} \d y \widetilde{\phi}_B(y)    \frac{(s^2+t^2+u^2)\;\Gamma(-ys)\Gamma(-yt)\Gamma(-yu)}
    {\Gamma(1+ys)\Gamma(1+yt)\Gamma(1+yu)},\hspace{-0.5cm}
\end{equation}
with $\widetilde{\phi}_B(y)=C_B \alpha'^4 (1-\alpha'/y)^3 (y/\alpha')^{-1}$, to the explicit amplitude \eqref{eq:fidModel}, we can significantly enhance the scattering strength, as shown in Figure \ref{fig:iu_addBG}. This particular background term is chosen because it generates a sizable imaginary part of the spin-0 partial wave near $s=3/(2\alpha')$. Establishing maximal analyticity for this block is substantially more involved than in the Veneziano case; the proof is given in Appendix~F.

\vspace{0.1cm}

\noindent{\bf Discussion:}
The construction presented in this paper suggests that the essential data of a unitary dual-resonance model need not be a discrete tower of zero-width states, but rather a spectral distribution over Regge slopes. In this language, ordinary dual models correspond to singular measures, while unitary dual amplitudes arise from sufficiently smooth distributions whose analytic structure supplies the absorptive cuts required by unitarity. The large-slope tail of the distribution is essential for partial-wave unitarity near threshold and at high spin. Singularities of the smearing function, on the other hand, determine the pattern of second-sheet resonances, whereas holomorphic weights generate nonresonant absorptive backgrounds. Thus, the smearing data provide a direct handle on both resonance physics and inelasticity.

We can further leverage this generic construction into a novel bootstrap program aimed at characterizing the Regge-slope spectral densities compatible with crossing, analyticity, locality, and, crucially, full unitarity. The examples studied here contain at most one family of second-sheet Regge trajectories, but more general smearing data can generate multiple trajectories and nonresonant absorptive backgrounds, with various different Regge cuts in the complex $\ell$ plane. Regge-slope smearing therefore provides a crossing-symmetric unitarization mechanism distinct from standard eikonalization, and opens an analytic route to classifying strongly coupled S-matrices that realize resonance--Regge duality beyond conventional string perturbation theory.
~\\

\noindent{\bf Acknowledgments:}
We would like to thank Shi-Lin Wan for helpful discussions. SYZ acknowledges support from the National Natural Science Foundation of China under Grant No.~12475074 and No.~12247103. The work of CdR and AJT is supported by STFC Consolidated Grant ST/X000575/1. CdR is also supported by a Simons Investigator award 690508. This research is also supported by the advanced computing resources provided by the Supercomputing Center of the USTC. CdR would like to thank the Perimeter Institute for Theoretical Physics for its hospitality while this work was finalized. \\

\bibliography{unitary_dual_model}

\appendix

\subsection{A. Unitarity violation in finitely smeared amplitudes}
\label{sec:uniViofinite}

In the main text, we pointed out that a finite-range smearing \cite{Martin:1969tc} of the form 
\begin{equation}
\label{eq:MansatzFinite}
    \mathcal M(s,t)
    =
    \!\int_{\alpha'}^{y_{\rm max}}\!\!\d y\, \phi(y)
    W_y(s,t,u)
\end{equation}
necessarily leads to unitarity violation in the high-spin partial waves.  To see this, let us examine the imaginary part of this amplitude in the $s$-channel, which receives contributions from the $V_y(s, t)$ and $V_y(s, u)$ blocks. The $tu$ crossing symmetry implies that contributions from $V_y(s, t)$ and $V_y(s, u)$ cancel each other for odd-$\ell$ partial waves, while for even-$\ell$ partial waves $V_y(s, t)$ and $V_y(s, u)$ give the same contribution. Using the fact that $\Im \big[1/(n-y (s+i0))]=\pi \delta(n-ys)$ and the expansion \eqref{eq:dualExpan}, we get, in the $s$-channel physical region ($s>0, ~t<0$), 
\begin{align}
\label{eq:ImMszpole}
\!\!\operatorname{Im} {\cal M}(s,t)&=\!\!\!\!
\sum_{ \lceil \alpha' s\rceil\leq n\leq \lfloor y_{\rm max} s\rfloor}\!\!\!\!
\phi\!\left(\frac{n}{s}\right)
\\&\times\bigg(\frac{\pi(\frac{n}{2}(z-1)+1)_n}{n!s} + \frac{\pi(\frac{n}{2}(-z-1)+1)_n}{n!s}\bigg), \nonumber
\end{align}
where we have also replaced $t$ in favor of $z=1+2t/s$. Since the Pochhammer symbol $(\frac{n}{2}(z-1)+1)_n$ is a polynomial in $z$, a finite $y_{\rm max}$ means that $\operatorname{Im} {\cal M}(s,t)$ is a polynomial with finite degree. Thus, there is a finite $\ell_{\rm max}$ beyond which $\operatorname{Im} a_{\ell>\ell_{\rm max}}(s)=0$. 
For the present model, however, $\Re a_\ell(s)\neq0$ for even $\ell>\ell_{\rm max}$, and therefore partial-wave unitarity is violated.

\subsection{B. Threshold behavior}

We now analyze the behavior of the amplitude near the threshold $s\to0$.  This is one of the asymptotic regions that cannot be fully tested by a finite numerical scan.  The large-$y$ tail of the smearing weight function controls the absorptive part of the amplitude in this limit. Reversing the argument, the absorptive part guides us in the design of acceptable weight functions. 

Let us assume that $\phi(y\to +\infty)\sim y^{-p}$, and use unitarity to constrain $p$. Near threshold, $\operatorname{Re}\mathcal M(s,t)$ is dominated by the massless pole of the Veneziano kernel: $W_y(s,t,u)\sim 1/s$, and thus
its partial wave is given by
\begin{equation}
    \operatorname{Re}a_\ell(s)
    \sim
    s^{-1+\frac{D-4}{2}} .
\end{equation}
To get the imaginary part, we can use the pole representation (Eq.~\eqref{eq:ImMszpole}), which, upon substituting $\phi(y\to +\infty)\sim y^{-p}$, leads to $\operatorname{Im}\mathcal M(s,t)\sim s^{p-1}$.
A partial-wave projection gives
\begin{equation}
    \operatorname{Im}a_\ell(s)
    \sim
    s^{p-1+\frac{D-4}{2}} .
\end{equation}
Then, partial-wave unitarity $|a_\ell(s)|^2\leq 2\,\operatorname{Im}a_\ell(s)$ requires $p\leq \frac{D-4}{2}-1$,
which in $D=10$ imposes the condition
\begin{equation}
    p\leq 2 .
\end{equation}

\subsection{C. Impact-parameter analysis}

We now analyze the large-energy and large-spin behavior of the smeared
amplitude.  This region is important because 
the massless pole gives a long-range tail that is most naturally studied in impact-parameter space.

We denote the transverse dimension by $d=D-2$. In the eikonal regime, the partial waves ($\ell\sim b\sqrt{s}/2$) can be approximated by the Fourier transform of the amplitude in the transverse momentum transfer:
\begin{equation}
\label{eq:absPartdef}
    a(b,s)
    =
    \frac{1}{2s}
    \int \frac{\d^d q}{(2\pi)^d}\,e^{i q\cdot b}
    \int_{\alpha'}^\infty \!\!\d y\,\phi(y)\,
    V_y(s,-q^2) ,
\end{equation}
where we have focused first on a single Veneziano channel.
The other crossed channels can be treated analogously and will be added in the final result. In the Regge limit (large $s$ and $q^2=-t\ll s$), using Stirling's formula, we can infer that the $y$-integral in Eq.~\eqref{eq:absPartdef} is localized in the region $yq^2\lesssim {\rm max}\{\frac{1}{\log(\alpha's)}, \alpha'q^2\}$, in which we have ${(yq^2)^2}/{ys}\ll1$. With the condition ${(yq^2)^2}/{ys}\ll1$, we can replace the Veneziano kernel in the integration by
\begin{equation}
    V_y(s,-q^2)
    \sim
    \Gamma(yq^2)\,
    e^{i\pi yq^2}\,
    \exp\!\left[-yq^2\log(ys)\right] .
\end{equation}
Since we are interested in the region $q^2\sim 0$ and $yq^2\sim 0$, we can approximate the prefactor with $    \Gamma(yq^2)e^{i\pi yq^2}= \frac{1}{yq^2}-\gamma+i\pi+\cdots$, where $\gamma$ is the Euler constant.
Thus, the leading imaginary part of the partial wave is
\begin{equation}
\label{eq:ImabsbFinal}
    \operatorname{Im}a(b,s)
    =
    \int_{\alpha'}^\infty \d y\,
    \frac{2^{-d-1}\pi^{1-d/2}\phi(y)}
    {s\,[y\log(ys)]^{d/2}}
    e^{-\frac{b^2}{4y\log(ys)}}.
\end{equation}
For the real part, making use of $e^{-\tau_0 q^2}/q^2=\int_{\tau_0}^\infty \d\tau\,e^{-\tau q^2}$, we get
\begin{equation}
    \operatorname{Re}a(b,s)
    =
    \int_{\alpha'}^\infty \d y\,
\frac{\pi^{-d/2}\phi(y)}
{2^{d+1}s\,y}
\int_{y\log(ys)}^\infty\!\!\!\!\!\!\!\!\! \d\tau\,
\tau^{-\frac{d}{2}} e^{-\frac{b^2}{4\tau}}.
\end{equation}

Let us now extract the asymptotic behavior in different impact-parameter regions.

{\bf Large impact parameter}
In the region $b^2\gg \alpha'\log(\alpha's)$,  Eq.~\eqref{eq:ImabsbFinal} is dominated by large values of $y$ for which the Regge width $y\log(ys)$ is comparable to $b^2$. So we can drop the integration below $y_\ast$: $y_\ast\log(sy_\ast)\equiv b^2$.
With $\phi(y)\sim y^{-p}$,
we can integrate to find that 
\begin{equation}
    \operatorname{Im}a(b,s)
    \sim
    \frac{1}{s}\,
    \frac{[\log(sb^2)]^{p-1}}
    {b^{d+2p-2}} .
\end{equation}
For the real part $\operatorname{Re}a(b,s)$, the $y$ integration above $y_*$ is subleading and can be dropped. For $y\ll y_\ast$, one has $b^2\gg y\log(ys)$, so we can approximate the $\tau$ integral with the Gamma function, which gives
\begin{equation}
    \operatorname{Re}a(b,s)
    \sim
    \frac{1}{s\,b^{d-2}}
    \int_{\alpha'}^{y_\ast} \d y\,\frac{\phi(y)}{y}\sim \frac{1}{s\,b^{d-2}} .
\end{equation}

Then, partial-wave unitarity $|a_\ell(s)|^2\leq 2\,\operatorname{Im}a_\ell(s)$ requires $2p-d+2\leq0$, which in $D=10$ gives $p\leq 3$. Note that this condition on $p$ is weaker than the condition extracted from the threshold, and thus can be dropped.

{\bf Crossover region} We now consider the region $b^2\lesssim \alpha' \log(\alpha's)$. In this region, the $y$-integrals of $\operatorname{Im}a(b,s)$ and $\operatorname{Re}a(b,s)$ are dominated by $y=O(\alpha')$, and the exponential factors are $O(1)$. Then, we can integrate to get
\begin{align}
    \operatorname{Im}a(b,s)
    &\sim
    \frac{1}{s[\log(\alpha's)]^{d/2}} ,\\
   \operatorname{Re}a(b,s)
    &\sim
    \frac{1}{s[\log(\alpha's)]^{d/2-1}} .
\end{align}
The unitarity ratio in this region behaves as
\begin{equation}
    \frac{|a(b,s)|^2}{2\operatorname{Im}a(b,s)}
    \sim
    \frac{1}{\alpha's}\,
    \frac{1}{[\log(\alpha's)]^{d/2-2}}  ,
\end{equation}
which decreases as $s$ increases. Thus, if unitarity is satisfied at finite $s$, we expect it to hold also at large $s$.

\subsection{D. Finite-energy, large-spin behavior}

We now look at the limit of  finite $s$ and large $\ell$, in which it is convenient to write $t=-s\varrho^2,~\varrho^2=({1-\cos\theta})/{2}$. In the $s$-channel, both $V_y(s,t)$ and $V_y(s,u)$ contribute to the absorptive part, but their partial-wave projections differ only by $(-1)^\ell$, since $t\leftrightarrow u$ sends $z\to-z$. Thus this only affects the even/odd projection and not the large-$\ell$ falloff. It is therefore enough to display the $V_y(s,t)$ block.

Our aim is to extract the leading non-analyticity at small $\varrho^2$, which controls the large-$\ell$ behavior. At fixed $s$, the $n$th term in the pole representation of the absorptive part probes the smearing function at $y=n/s$. If the smearing range had a finite upper endpoint, only finitely many $n$ would contribute, and the resulting dependence on $\varrho^2$ would be analytic. Thus, in the present model, the non-analyticity can only originate from the unbounded large-slope tail. Keeping only the leading large-$n$ asymptotics relevant for this non-analyticity, the sum takes the form
$
    S_p(\varrho^2)
    =
    \sum_{n\geq \lceil\alpha's\rceil}
    n^{-p}e^{-\varrho^2n\log n} 
$. For $p>1$, $S_p(0)$ is finite. The coefficient of $(\varrho^2)^k$ in its formal Taylor expansion is proportional to $\sum n^{k-p}(\log n)^k$, which converges precisely when $k<p-1$. The convergent coefficients receive their main contributions from a finite range of $n$, corresponding at fixed $s$ to a finite range of Regge slopes, and give analytic terms in $\varrho^2$. By contrast, as $\varrho\to0$, the contributions associated with the divergent Taylor coefficients are controlled by increasingly large $n$, and hence probe increasingly large Regge slopes. It is this growing sensitivity to the large-$n$ tail that prevents the small-$\varrho^2$ expansion from being interchanged with the sum and generates the non-analyticity.

Let $m$ denote the highest Taylor order with a convergent coefficient. Explicitly, $m=\lfloor p-1\rfloor$ for $p\notin\mathbb Z$, whereas $m=p-2$ for $p\in\mathbb Z$. We then subtract all Taylor terms through order $m$ and define the remainder
\begin{equation}
    \!\!\!R_p(\varrho^2)
    =\!\!\!\!\!
    \sum_{n\geq \lceil\alpha's\rceil}\!\!\!\! n^{-p}
    \left[
    e^{-\varrho^2n\log n}
    -\!\!
    \sum_{k=0}^{m}
    \frac{(-\varrho^2n\log n)^k}{k!}
    \right]\!\!.
\end{equation}
To estimate this remainder, we introduce $n_\ast$ through $\varrho^2n_\ast\log n_\ast\simeq1$.

If $p$ is not an integer, then $R_<\!\sim\! (\varrho^2)^{m+1}n_\ast^{m+2-p}(\log n_\ast)^{m+1}$, while $R_>\sim (\varrho^2)^m n_\ast^{m+1-p}(\log n_\ast)^m$, and both scale as
\begin{equation}
    R_p(\varrho^2)\sim
    (\varrho^2)^{p-1}
    \left(\log {1\over \varrho^2}\right)^{p-1}, ~ p\notin\mathbb Z.
\end{equation}

If $p$ is an integer, so that $p=m+2$, the first non-subtracted term is marginal:
$R_<\sim (\varrho^2)^{m+1}\sum_{n<n_\ast}(\log n)^{m+1}/n\sim (\varrho^2)^{m+1}(\log(1/\varrho^2))^{m+2}$, while $R_>\sim (\varrho^2)^{m+1}(\log(1/\varrho^2))^{m+1}$ is subleading. Thus
\begin{equation}
   R_p(\varrho^2)
    \sim
    \varrho^{2p-2}
    \left(\log\frac{1}{\varrho^2}\right)^p,
    ~ p\in\mathbb Z .
\end{equation}

The extra logarithm at integer $p$ does not change the large-$\ell$ scaling. In the large-$\ell$ limit, the partial-wave projection is equivalent to a transverse Fourier transform with $b\simeq 2\ell/\sqrt{s}$,
\begin{equation}
    \!\!\!\int \frac{\d^d q}{(2\pi)^d}\, e^{iq\cdot b}(q^2)^\alpha
    =
    C(\alpha)b^{-d-2\alpha},
    ~
    C(\alpha)\propto {1\over\Gamma(-\alpha)} .
\end{equation}
For the non-integer case, this directly gives $(\log b)^{p-1}/b^{d+2p-2}$. On the other hand, for integer $p=m+2$, the relevant power is $\alpha=p-1=m+1$, for which $C(m+1)=0$; differentiating $m+2$ times with respect to $\alpha$ removes the would-be $(\log b)^{m+2}$ term and leaves $(\log b)^{m+1}/b^{d+2(m+1)}$.
Therefore, in both cases,
\begin{equation}
    \operatorname{Im}a_\ell(s)
    \sim
    {(\log\ell)^{p-1}\over \ell^{d+2p-2}} \,,
\end{equation}
which agrees with the large-impact-parameter estimate derived in the eikonal regime.

The real part of the amplitude is controlled by the massless pole at $t=0$, which gives, in the large $\ell$ limit,
\begin{equation}
    \operatorname{Re}a_\ell(s)
    \sim
    \frac{1}{\ell^{d-2}} .
\end{equation}
Consequently, partial-wave unitarity in this case requires
\begin{equation}
    2p-d+2\leq0,
\end{equation}
which in $D=10$ gives $p\leq3$.

\subsection{E. Analyticity and locality via explicit continuation}

In the main text, we have proven that our dual amplitudes satisfy Mandelstam's maximal analyticity using the Euler-Beta representation and a Landau analysis. Here, we provide an alternative proof of analyticity for general $s$ and $t<0$, as well as locality. That is, we show that our dual amplitudes satisfy the fixed-$t$ dispersion relation. 

At fixed $t$, the real-$y$ integration defining our unitary dual amplitude ${\cal M}(s,t)$ converges in most of the complex $s$ plane, except within a small region surrounding the segment $0<s<-t$. The precise shape of this non-convergent region depends mildly on the block under consideration and is controlled by the large positive-$y$ asymptotics
\begin{align}
|V_y(s, u)| &\sim y^{-\frac{1}{2}}  e^{y \Re [-s\log (-\alpha's) -t\log(-\alpha' t)-u\log(-\alpha' u) ]},  
\nonumber\\
|V_y(s, t)| &\sim y^{-\frac12} e^{y \Re [-s\log (-\alpha's) -t\log(-\alpha't) -u\log(\alpha'u) ]},\nonumber
\end{align}
with $|V_y(u, t)|$'s region the same as $|V_y(s, t)|$'s. Here the estimate for $V_y(s,u)$ keeps only the large-$y$ exponential rate for $y>0$, while the full asymptotics also contain the reflection-formula factor $\sin(\pi y t)$, which is bounded for fixed real $t<0$ and hence does not affect the convergence of the real-$y$ integral.
Whenever the real part of the exponent is positive, the integrand grows exponentially at large $y$, and the corresponding $y$ integral becomes divergent.

However, this small region is also analytic away from the real $s$ axis. In the following, we will show this for the ${\cal V}(s,t)$ (or ${\cal V}(u,t)$) and ${\cal V}(s,u)$ blocks separately, by analytically continuing and verifying the fixed $t$ dispersion relation. 

$\bullet$ For the ${\cal V}(s,t)$ (or similarly ${\cal V}(u,t)$) block, the analytic continuation can be implemented by simply rotating the $y$-integration contour away from the real-$y$ axis (see Figure \ref{fig:iu_deform_y}), with the residue contributions from any poles crossed during the deformation included. One can verify numerically that, at fixed $t$, as the rotation angle $\varphi_y$ increases, the analyticity domain of ${\cal V}(s,t)|_{\varphi_y}$ expands to cover an increasing portion of the small region surrounding the segment $0<s<-t$, namely the nonanalytic region of ${\cal V}(s,t)|_{\varphi_y=0}$. Above a critical value $\varphi^c_y(t)$, this analyticity domain covers the upper half of the nonanalytic region of ${\cal V}(s,t)|_{\varphi_y=0}$ and at the same time overlaps with the analytic region of ${\cal V}(s,t)|_{\varphi_y=0}$, thus constituting an analytic continuation. 
(The analytic continuation can also be done via the Abel-Pad\'e method that will be discussed shortly, and we have checked that the two methods agree with each other.)

With the analytic continuation established, we can verify that the ${\cal V}(s,t)$ block indeed satisfies an unsubtracted dispersion relation
\begin{equation}
    \bar{{\cal V}}(s, t) = \frac{1}{\pi}\int_{0}^{\infty} \d \mu\, \frac{\Im \bar{{\cal V}}(\mu+i0, t)}{\mu-s}\, ,
\end{equation}
where for clarity we have subtracted the $s=0$ poles on both sides, $\bar{{\cal V}}(s, t)={\cal V}(s, t)-(...)/s$. For example, choosing six representative points in the $s,t$ plane and taking $\phi (y) = 16\alpha'^2(1-\alpha'/y)/(y/\alpha')^2$, we find that the dispersion relation indeed reconstructs our original dual amplitude block:

{\small
\begin{equation}
\begin{array}{c|c|c}
    (\alpha' s,~\alpha' t) & 
\bar{\mathcal V}_{\rm disp}(s,t)/(2\alpha')^3
&
\bar{\mathcal V}_{\rm direct}(s,t)/(2\alpha')^3\\
    \hline
    (-47/14,~ -1/4) & 0.340 & 0.340\\
    (-i/8, ~-1/4) & 0.635+0.071i & 0.635+0.071i \\
    (3/2+i/2, ~-1/4) & 0.357+0.424i & 0.357+0.424i \\
    (2+3i/4, ~-1/4) & 0.302+0.378i  & 0.302+0.378i \\
    (-11/14,~ -1/6)  & 1.049  & 1.049\\
    (1/2+i/2,~-1/6) & 1.181+0.589i & 1.181+0.589i    
\end{array}\nonumber
\end{equation}
}

$\bullet$ For the ${\cal V}(s,u)$ block, however, the same contour rotation does not work. This is because, in the $0<s<-t$ region, we have $u>0$, and thus the $s$- and $u$-channel poles both lie along the positive $y$ axis, with the $s$-channel poles below the axis and the $u$-channel poles above it. This is different from Figure \ref{fig:iu_deform_y}, which pinches the $y$ contour to the real axis. Thus, in that case, we instead use an Abel-summed pole representation, and then evaluate the continuation numerically by a Pad\'e approximation. This gives the discontinuity within the interval $0<s<-t$ needed for the dispersion-relation check.

Recall that the $s$-channel discontinuity of ${\cal V}(s, u)$ can be written as
\begin{align}
    {\rm Disc}_s {\cal V}(s, u) &\equiv \frac{1}{2i}\big[{\cal V}(s+i0, u-i0)-{\cal V}(s-i0, u+i0)\big]\, \nonumber
    \\
    &=\sum_{n\geq \lceil \alpha' s\rceil} c_n(s, t) - \sum_{n\geq \lceil \alpha' u\rceil} c_n(u, t)\,,\nonumber
\end{align}
where $c_n(s, t) = (-1)^n \frac{\pi}{s} \frac{\phi (n/s) \Gamma(n(1+t/s))}{\Gamma(n+1)\Gamma(n(t/s))}$. For $0<s<-t$, this sum is divergent and must be understood by analytic continuation.  At large $n$, $c_n(s, t)\sim e^{H(s, t) n/s}$, where $H(s, t)=-s\log(\alpha's)-t\log(-\alpha' t) -u\log(\alpha'u)>0$ in the interval $0<s<-t$. 
This motivates us to introduce the Abel regulated discontinuity
\begin{equation}
    {\rm Disc}_s {\cal V}^{(\epsilon)}\!(s, u) :=\!\!\!\!\sum_{n\geq \lceil \alpha' s\rceil}\!\!\!\! c_n(s, t) e^{-\epsilon \frac{n}{s}} - \!\!\!\!\sum_{n\geq \lceil \alpha' u\rceil}\!\!\!\! c_n(u, t)e^{-\epsilon \frac{n}{u}} \,. \nonumber
\end{equation}
This sum converges if we choose $\epsilon>H$. The next step is to construct a Pad\'e approximation in the convergent region of $\epsilon$ and analytically continue it to $\epsilon=0$. To this end, we Taylor-expand ${\rm Disc}_s {\cal V}^{(\epsilon)}\!(s, u)$ around, say, $\epsilon=H+\frac{1}{4\alpha'}$ to a certain order, and then approximate this truncated Taylor series with a diagonal Pad\'e series, i.e., a rational function whose numerator and denominator are polynomials of equal order. Then we take the limit $\epsilon=0$, which completes the Abel-Pad\'e continuation if numerical stability is observed as the order of the Pad\'e approximant is increased. 

\begin{figure}[htbp]
     \centering
     \includegraphics[width=0.8\linewidth]{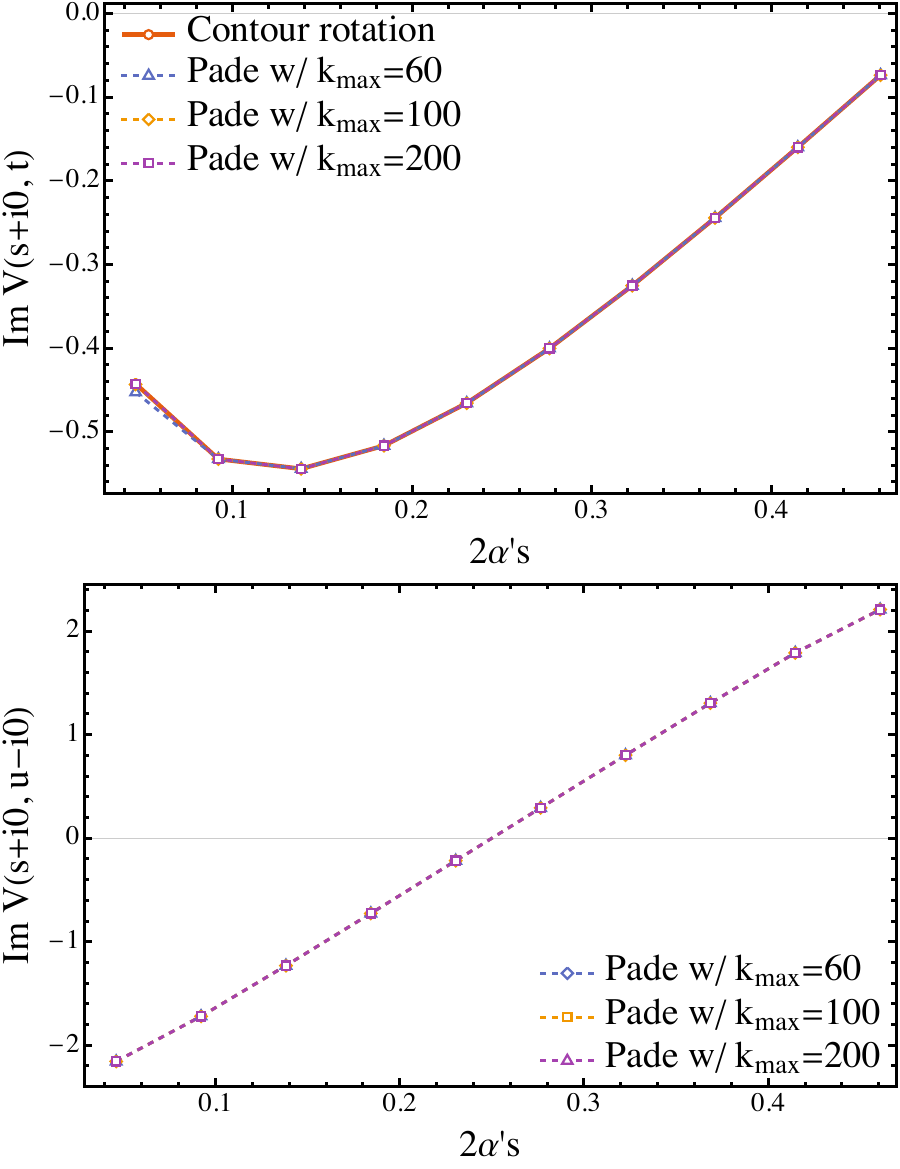}
     \caption{
     Analytic continuation at $-\alpha't=1/4$. The \textit{Top} panel shows that, for the $\mathcal{V}(s,t)$ block, the contour rotation method agrees with the Abel-Pad\'e method. The \textit{Bottom} panel demonstrates the validity of the Abel-Pad\'e continuation method for $\mathcal{V}(s,u)$ with different Pad\'e orders.}
     \label{fig:iu_continue}
\end{figure}

Using this continued discontinuity, we reconstruct the ${\cal V}(s,u)$ block from the fixed-$t$ dispersion relation
\begin{equation}
    \!\bar{{\cal V}} (s, u) = \!\int_{-t/2}^{\infty}\!\! \frac{\d \mu}{\pi} \!\left[\frac{{\rm Disc}_s \bar{{\cal V}}(\mu, -\mu-t)}{\mu-s} +(s\to u)\right]\,,
\end{equation}
where the barred quantity $\bar{{\cal V}} (s, u)$ again denotes the amplitude with the $s=0$ and $u=0$ poles subtracted.
We can numerically verify that ${\cal V} (s, u)$ obtained in this way agrees with the result from the $y$ integration. 
For example, with $\phi(y) = 16\alpha'^2 (1-\alpha'/y)/(y/\alpha')^2$, we have, after subtracting the $s,~u=0$ poles,

{\small
\begin{equation}
\begin{array}{c|c|c}
    (\alpha' s,~\alpha' t) &
\frac{\bar{\mathcal V}_{\rm disp}(s,u)}{(2\alpha')^3}
&
\frac{\bar{\mathcal V}_{\rm direct}(s,u)}{(2\alpha')^3}\\
    \hline
    (i/4,~ -1/4)  & 0.496-0.287i  & 0.496-0.287i\\
    (101/200+i/2, ~-1/4) & 0.099+0.176i&  0.099+0.176i \\
    (3/2+i/2, ~-1/4) & -0.033+0.019i&  -0.033+0.019i \\
    (2+3i/4, ~-1/4)& -0.014+0.014i&  -0.014+0.014i \\
    (-51/200 + i/20,~ -1/6)  & 0.292+0.767i  & 0.292+0.767i\\
\end{array}\nonumber
\end{equation}
}

Thus, we have established that our dual amplitude satisfies an unsubtracted dispersion relation at fixed $t<0$.

\subsection{ F. Continuation for the Virasoro-Shapiro-like block}

In the main text, we have also used the Virasoro-Shapiro-like block as an absorptive background to enhance the inelasticity or coupling strength. In this appendix, we show that, as in the Veneziano case, this Virasoro-Shapiro-like block satisfies Mandelstam's maximal analyticity condition, namely that it only has branch cuts for
\begin{equation}
    s\in[0,\infty)~~~\text{or}~~~ t\in[0,\infty)~~~\text{or}~~~ u\in[0,\infty).
\end{equation}

As in the Veneziano case, we begin with the complex Beta integral
\begin{equation}\label{eq:complex_beta_1}
I_y(a,b)
=
\int_{\mathbb C}\d^2 \zeta\,
|\zeta|^{-2ya-2}
|1-\zeta|^{-2yb-2}.
\end{equation}
By analytic continuation of the standard complex Beta integral, one has
\begin{equation}\label{eq:complex_beta_2}
I_y(a,b)
=
\pi
\frac{
\Gamma(-ya)\Gamma(-yb)\Gamma(1+y(a+b))
}{
\Gamma(1+ya)\Gamma(1+yb)\Gamma(-y(a+b))
},
\end{equation}
which leads to
\begin{align}
\label{eq:IyVS}
&~~~~ -\frac{1}{\pi y^2}
\left[I_y(s,t)+I_y(t,u)+I_y(u,s)
\right]\nonumber\\
&=\frac{(s^2+t^2+u^2)\;\Gamma(-ys)\Gamma(-yt)\Gamma(-yu)}
    {\Gamma(1+ys)\Gamma(1+yt)\Gamma(1+yu)}.
\end{align}
We want to determine the analyticity of Eq.~\eqref{eq:IyVS} upon integration over $y$. For this, it suffices to analyze one term in this cyclic sum. 

Let us define 
\begin{align}
     \mathcal{I}(s,t)
     &=
     -\frac{1}{\pi}
     \int_{\alpha'}^{\infty}\d y\,
     \frac{\widetilde{\phi}_B(y)}{y^2}
     I_y(s,t)
     \nonumber\\
     &=
     \int_{\mathcal{C}}
     \frac{\d^2\zeta}{|\zeta|^2|1-\zeta|^2}
     G_\phi\!\left(\mathcal{A}_{s,t}(\zeta,\bar\zeta)\right),
     \label{eq:exchange_VS_block}
\end{align}
where the second equality is understood formally at this stage and will be made precise below by complexifying $\bar\zeta$ and specifying the contour ${\cal C}$, and
\begin{align}
    G_\phi(\mathcal{A})
    &=
    -\frac{1}{\pi}
    \int_{\alpha'}^{\infty}\d y\,
    \frac{\widetilde\phi_B(y)}{y^2}
    e^{-\mathcal{A}y},
\\
    \mathcal{A}_{s,t}(\zeta,\bar\zeta)
    &=
    s\log|\zeta|^2+t\log|1-\zeta|^2.
\end{align}
Unlike for the Veneziano block, where $\operatorname{Re}s<0$ and $\operatorname{Re}t<0$ provide an undeformed contour with $\operatorname{Re}\mathcal A_{s,t}>0$, the second equality in Eq.~\eqref{eq:exchange_VS_block} does not follow here from a common undeformed convergence region valid for all $y$. For our purpose, it is sufficient to justify the exchange in a neighbourhood of a single suitable base point, thereby identifying the two representations as the same analytic germ. The required contour will be constructed later, at the base point introduced in Eq.~\eqref{eq:base_point}.

As in the main text, $G_\phi(\mathcal{A})$ can be continued to an analytic function of $\mathcal{A}$ except for a finite branch point at $\mathcal{A}=0$. Also, there are three degenerate cases of the integration, in which $|\mathcal{A}_{s,t}(\zeta,\bar\zeta)|\to\infty$ for $s,~t,~u\neq0$,
\begin{equation}
    \zeta=0,\qquad \zeta=1,\qquad \zeta=\infty.
\end{equation}

Note that the degenerate cases can still be well-defined as long as $G_\phi(\mathcal{A})$ decays exponentially as $\mathcal{A}\to\infty$. This happens in any sector with $\Re\mathcal{A}>0$ on the first Riemann sheet. We will show that, by contour deformation, all the degenerate cases fall in this class. For this, let us uplift $\bar\zeta$ to an independent complex variable, and the original integration contour lies in the subspace of the 2D complex space $\mathbb{C}^2$:
\begin{equation}
    \mathcal{C}_E=\{(\zeta,\widetilde{\zeta}):\widetilde{\zeta}=\bar\zeta\} .
\end{equation}
We want to deform the contour within the 2D complex space, making the replacements
\begin{equation}\label{eq:complexify_zeta}
    |\zeta|^2\to \zeta\widetilde{\zeta},
    \qquad
    |1-\zeta|^2\to (1-\zeta)(1-\widetilde{\zeta})\,.
\end{equation}
Here and below, each deformed contour is regarded as an oriented real two-dimensional contour, with integration measure given by the holomorphic two-form
\begin{equation}
    \Omega\equiv
    \frac{i}{2}
    \frac{\d\zeta\wedge\d\widetilde\zeta}
    {\zeta\widetilde\zeta(1-\zeta)(1-\widetilde\zeta)},
\end{equation}
whose restriction to $\mathcal C_E$ is the original measure $\d^2\zeta/(|\zeta|^2|1-\zeta|^2)$.
Near $\zeta=0$, the original contour can be parametrized as
\begin{equation}\label{eq:VS_origin_contour}
   \! \zeta=\zeta_0e^{-R+i\theta},\,
    \widetilde{\zeta}=\widetilde{\zeta}_0 e^{-R-i\theta}\!,
    ~
    R\to+\infty,\theta\in[0,2\pi),
\end{equation}
and we can deform the radial direction by taking
\begin{equation}\label{eq:choose_vs}
    \zeta=\zeta_0 e^{-R v_s+i\theta},
    \qquad
    \widetilde{\zeta}=\widetilde{\zeta}_0 e^{-R v_s-i\theta}.
\end{equation}
The condition that the deformed contour still ends at
$\zeta=\widetilde{\zeta}=0$ is simply $\Re v_s>0$.
Near this endpoint, $\mathcal{A}_{s,t}(\zeta,\widetilde{\zeta})
    \sim
    s\log \zeta\widetilde{\zeta} $,
and hence the endpoint integral converges if $ \Re(-s v_s)>0$. Such a direction $v_s$ exists precisely when
\begin{equation}
    s\notin[0,\infty).
\end{equation}

The endpoint $\zeta=1$ can be treated in the same way, using local coordinates $1-\zeta$ and $1-\widetilde{\zeta}$. The corresponding contour deformation is unobstructed precisely when
\begin{equation}
    t\notin[0,\infty).
\end{equation}
For the endpoint \(\zeta=\infty\), we set
\begin{equation}
    \zeta=\frac{1}{\chi},
    \qquad
    \widetilde{\zeta}=\frac{1}{\widetilde{\chi}},
\end{equation}
and consider \(\chi,\widetilde{\chi}\to0\). In this region, $\mathcal{A}_{s,t}(\chi,\widetilde{\chi})
    \sim
    -(s+t)\log(\chi\widetilde{\chi})$.
Thus, the same endpoint analysis shows that the deformation is unobstructed precisely when
\begin{equation}
    -(s+t)\notin[0,\infty),~~~{\rm or}~~~u\notin[0,\infty) .
\end{equation}

It remains to exclude possible internal pinch singularities. Since $G_\phi(\mathcal A)$ has a finite limiting value at $\mathcal A=0$, any simple preimage of $\mathcal A_{s,t}=0$ can be bypassed by an arbitrarily small local detour. Simple intersections with $\mathcal A_{s,t}=0$ are therefore harmless; an internal obstruction can arise only from a critical encounter of this branch locus with the chosen contour, whose necessary condition is given by the Landau equations
\begin{equation}
    \mathcal A_{s,t}(\zeta_*,\widetilde{\zeta}_*)=0,
    ~
    \partial_\zeta\mathcal A_{s,t}(\zeta_*,\widetilde{\zeta}_*)=0,
    ~
    \partial_{\widetilde{\zeta}}\mathcal A_{s,t}(\zeta_*,\widetilde{\zeta}_*)=0 .
\end{equation}
Throughout $\mathcal D_{s,t}$, the variables $s,t,u$ are nonzero. Hence $s+t=-u\neq0$, and the last two Landau equations imply
\begin{equation}
 \zeta_*=\widetilde{\zeta}_*=\frac{s}{s+t}\equiv r.
\end{equation}
For each fixed $(s,t)\in\mathcal D_{s,t}$, the quantity $r$ is finite and distinct from $0$ and $1$. The endpoint deformations can therefore be confined to sufficiently small neighborhoods of $\zeta=0,1,\infty$ that avoid $r$, while the complementary finite part is kept on the initial contour $\widetilde{\zeta}=\bar\zeta$; this critical point can pinch the latter only when $r$ is real.

For finite real $r\neq0,1$, substituting the critical point into
$\mathcal A_{s,t}=0$ gives
\begin{equation}
    r\log(r^2)+(1-r)\log\!\big((1-r)^2\big)=0 .
\end{equation}
This equation has no finite real solution: its left-hand side
is negative for $0<r<1$ and positive for $r<0$ or $r>1$.
Therefore there is no additional internal first-sheet pinch.

We finally establish that the analytic continuation is independent of the path in $(s,t)$ space. The preceding endpoint and no-pinch analysis ensures local continuation along every path in
\begin{equation}
    \mathcal D_{s,t}
    =
    \left\{
    (s,t)\in\mathbb C^2
    \,\middle|\,
    s,\ t,\ u=-s-t\notin[0,\infty)
    \right\}.
\end{equation}
It remains to exclude a possible global monodromy. Unlike a single Veneziano block in the main text, whose continuation domain is the simply connected product of two slit planes, the present domain also imposes $u=-s-t\notin[0,\infty)$ and contains one noncontractible cycle. To see this, write $\nu=\xi_\nu+i \eta_\nu$ for $\nu\in\{s,t,u\}$ and define $P^\times=\{\boldsymbol \eta\in\mathbb R^3:\eta_s+\eta_t+\eta_u=0,\ \boldsymbol \eta\neq0\}$. The origin is excluded because $\eta_s=\eta_t=\eta_u=0$ would require $\xi_s,\xi_t,\xi_u<0$, contradicting $\xi_s+\xi_t+\xi_u=0$. For each $\boldsymbol \eta\in P^\times$, the allowed real parts form a convex set. A continuous choice within this set is
\begin{equation}
    \xi_\nu^\star(\boldsymbol \eta)
    =
    \frac{\eta_\nu^2-|\boldsymbol \eta|^2/3}{|\boldsymbol \eta|},
    \qquad
    \nu\in\{s,t,u\},
\end{equation}
for which $\xi_\nu^\star=-|\boldsymbol \eta|/3<0$ whenever $\eta_\nu=0$. Linear interpolation to this choice gives a strong deformation retraction, so
\begin{equation}
    \mathcal D_{s,t}
    \simeq
    P^\times
    \cong
    \mathbb R^2\setminus\{0\}
    \simeq
    S^1,
    \qquad
    \pi_1(\mathcal D_{s,t})\cong\mathbb Z.
\end{equation}

We may therefore check the monodromy along a single generating cycle. For $\kappa>0$, we define
\begin{align}
    c_k(\theta)
    &=
    \cos\left(\theta+\frac{2\pi k}{3}\right), \nonumber
    \\
    \omega_k(\theta)
    &=
    \kappa\left(c_k(\theta)^2-\frac12+i c_k(\theta)\right),
    ~
    k=0,1,2,
\end{align}
and set $(s,t,u)=(\omega_0,\omega_1,\omega_2)$ for $0\leq\theta\leq2\pi$. The identities $\sum_kc_k=0$ and $\sum_kc_k^2=3/2$ imply $s+t+u=0$. Moreover, if $c_k\neq0$, then $\omega_k$ is nonreal, while if $c_k=0$, then $\omega_k=-\kappa/2<0$. The cycle therefore remains in $\mathcal D_{s,t}$, and its imaginary part winds once around the origin of $P^\times$, so it represents a generator of $\pi_1(\mathcal D_{s,t})$. At $\theta=0$, its base point is
\begin{equation} \label{eq:base_point}
    (s,t,u)
    =
    \kappa\left(
    \frac12+i,
    -\frac14-\frac{i}{2},
    -\frac14-\frac{i}{2}
    \right),~~\kappa>0
\end{equation}
which lies in the starting region used above.

Along this cycle it is convenient to use the fixed-real-$y$ representation
\begin{align}
    \mathcal{I}(s,t)&=u^2\mathcal{J}(s,t,u),
    \\
    \mathcal{J}(s,t,u)
    &=
    \int_{\alpha'}^\infty \d y\,\widetilde\phi_B(y)
    \prod_{\nu\in\{s,t,u\}}
    \frac{\Gamma(-y\nu)}{\Gamma(1+y\nu)}. \nonumber
\end{align}
For fixed $y>0$, the integrand is single-valued and holomorphic on $\mathcal D_{s,t}$. Using the reflection formula together with the uniform Stirling expansion gives
\begin{equation}
    \bigg|\prod_{\nu\in\{s,t,u\}}
    \frac{\Gamma(-y\nu)}{\Gamma(1+y\nu)}\bigg|
    = \mathcal{O}\left( y^{-3}e^{-y\mathcal{E}(s,t,u)}\right),
\end{equation}
where $\mathcal E(s,t,u)$ is the real exponential rate controlling the large-$y$ magnitude of the integrand. We have
\begin{equation}
    \mathcal{E}(s,t,u)
    =
    2\operatorname{Re}\!\!\!\!
    \sum_{\nu\in\{s,t,u\}}
    \!\!\!\nu\log(-\alpha'\nu)
    \!-\!
    \pi
    \!\!\!\!\sum_{\nu\in\{s,t,u\}}
    \!\!\!|\operatorname{Im}\nu|.
\end{equation}
Direct substitution of the generating cycle gives
\begin{equation}
    \mathcal{E}\bigl(
    \omega_0(\theta),
    \omega_1(\theta),
    \omega_2(\theta)
    \bigr)
    \geq
    \kappa\log2>0\,.
\end{equation}
Since the cycle is compact, the same estimate, with some positive lower bound $\delta$, holds in an open tubular neighbourhood $U\subset\mathcal D_{s,t}$ of the entire cycle. As $\widetilde\phi_B(y)=O(y^{-1})$ for large $y$, while its behaviour at $y=\alpha'$ is integrable, one has
\begin{equation}
    \bigg|\widetilde\phi_B(y)\prod_{\nu\in\{s,t,u\}}
    \frac{\Gamma(-y\nu)}{\Gamma(1+y\nu)}\bigg|
    \leq
    C y^{-4}e^{-\delta y}
\end{equation}
uniformly on compact subsets of $U$. 
The bound above ensures that the $y$ integral converges locally uniformly on $U$. Since the integrand is single-valued and holomorphic in $(s,t)$ for every fixed $y$, $\mathcal{J}_U$ is a single-valued holomorphic function on $U$. Every closed path in $\mathcal D_{s,t}$ is homotopic to an integer power of the generating cycle, so it is sufficient to prove that the continuation around this cycle has trivial monodromy. Since $u^2\mathcal{J}_U$ is already single-valued in a neighbourhood of the cycle, this will follow once we show that, at the base point, it defines the same analytic germ as the worldsheet representation in Eq.~\eqref{eq:exchange_VS_block}.

The discussion above leaves only one point to prove: the single-valued function $u^2\mathcal{J}_U$ must agree, near the base point Eq.~\eqref{eq:base_point}, with the germ defined by the worldsheet representation in Eq.~\eqref{eq:exchange_VS_block}. We now construct the worldsheet contour required for this identification. The identification proceeds in two steps. We first construct from the endpoint-regularized contour a $y$-independent contour $\mathcal C_+$ allowing the exchange of the $y$ and worldsheet integrations, and then show that $\mathcal C_+$ gives the same $G_\phi$ integral as the original endpoint-deformed contour by a non-pinched lifted homotopy.

Let $(s_0,t_0,u_0)$ denote the base point in Eq.~\eqref{eq:base_point}, and write $\mathcal A_0=\mathcal A_{s_0,t_0}$. On the original surface $\mathcal C_E$,
\begin{equation}
    \mathcal A_0
    =
    s_0\left(
        \log|\zeta|^2
        -
        \frac{1}{2}\log|1-\zeta|^2
    \right).
\end{equation}
The expression in parentheses is real. Consequently, $\operatorname{Re}\mathcal A_0\to-\infty$ only at the endpoint $\zeta=0$, whereas it tends to $+\infty$ at $\zeta=1$ and $\zeta=\infty$. We therefore deform only the $\zeta=0$ end. More explicitly, for sufficiently large $R_0$, the part with $R\geq R_0$ in Eq.~\eqref{eq:VS_origin_contour} may be replaced by
\begin{equation}
    \zeta
    =
    \zeta_0
    e^{-R_0-(R-R_0)e^{i\pi/4}+i\theta},
    ~
    \widetilde\zeta
    =
    \widetilde\zeta_0
    e^{-R_0-(R-R_0)e^{i\pi/4}-i\theta}.
\end{equation}
This corresponds to the choice $v_{s_0}=e^{i\pi/4}$ in Eq.~\eqref{eq:choose_vs}. Along this contour,
\begin{equation}
    \mathcal A_0
    =
    -2s_0e^{i\pi/4}R+\mathcal{O}(1)
    =
    \frac{\kappa R}{\sqrt{2}}(1-3i)+\mathcal{O}(1).
\end{equation}
Thus, by taking $R_0$ sufficiently large, the entire deformed tail can be chosen so that
\begin{equation}
    \operatorname{Re}\mathcal A_0\to +\infty,
    \qquad
    \operatorname{Im}\mathcal A_0<0.
\end{equation}
We denote the resulting integration surface by $\mathcal C_{\rm reg}$. For every fixed $y>0$, after fixing the logarithmic lift inherited from $\mathcal C_E$, the surface $\mathcal C_{\rm reg}$ is obtained by continuously deforming the original integration surface along a path in parameter space from a point where the complex Beta integral is absolutely convergent to the base point, while keeping its noncompact ends in the corresponding decay sectors. Its fixed-$y$ integral therefore gives the analytic continuation of $I_y(s,t)$ and remains equal to Eq.~\eqref{eq:complex_beta_2}.

This endpoint regularization is sufficient for each fixed-$y$ integral, but it does not yet justify exchanging the $y$ and $(\zeta,\widetilde{\zeta})$ integrations: $\operatorname{Re}\mathcal A_0$ may still be nonpositive on a compact middle part of $\mathcal C_{\rm reg}$. We shall continuously deform $\mathcal C_{\rm reg}$ to a $y$-independent surface on which $\operatorname{Re}\mathcal A_0$ has a uniform positive lower bound.

For this purpose, follow the logarithms continuously so that
$\mathcal A_0$ is single-valued on the corresponding logarithmic
covering space, and set $\mathcal R=\operatorname{Re}\mathcal A_0$. We equip this covering space with the pullback of the complete Hermitian metric
\begin{equation}
    g
    =
    \left(
        \frac{1}{|\zeta|^2}
        +
        \frac{1}{|1-\zeta|^2}
    \right)|d\zeta|^2
    +
    \left(
        \frac{1}{|\widetilde\zeta|^2}
        +
        \frac{1}{|1-\widetilde\zeta|^2}
    \right)|d\widetilde\zeta|^2.
\end{equation}
All gradients and norms below are taken with respect to $g$. Define the upward flow $\Phi_\tau$ by
\begin{equation}
    \frac{\d}{\d\tau}\Phi_\tau(X)
    =
    \left.
    \frac{\nabla_g\mathcal R}
    {1+\|\nabla_g\mathcal R\|_g^2}
    \right|_{\Phi_\tau(X)},
    \qquad
    \Phi_0=\operatorname{id}.
\end{equation}
The denominator bounds the $g$-speed without changing the flow
trajectories. Together with the completeness of $g$, this makes the finite-time flow well defined, as verified below. Since $\mathcal A_0$ is holomorphic and $g$ is Hermitian, the Cauchy--Riemann equations give
\begin{align}
    \frac{\d}{\d\tau}
    \mathcal R(\Phi_\tau(X))
    &=
    \frac{
        \|\nabla_g\mathcal R(\Phi_\tau(X))\|_g^2
    }{
        1+\|\nabla_g\mathcal R(\Phi_\tau(X))\|_g^2
    }
    \geq0, \nonumber
    \\
    \frac{\d}{\d\tau}
    \operatorname{Im}&\mathcal A_0(\Phi_\tau(X))
    =
    0.
\end{align}
The real part therefore increases along the flow, while the imaginary part remains fixed.

We now show that the flow moves every point of the integration surface above a fixed positive level. By a saddle we mean a critical point of
$\mathcal A_0$,
\begin{equation}
    \d\mathcal A_0=0,
\end{equation}
or equivalently, by the Cauchy--Riemann equations, $\nabla_g\mathcal R=0$. A trajectory that failed to reach the desired positive region would have two apparent possibilities: it could accumulate at a saddle with $\mathcal{R}<0$, or it could enter a noncompact region while $\mathcal R$ remains negative. Such noncompact regions arise when either $\zeta$ or $\widetilde\zeta$ approaches $0$, $1$, or $\infty$,
or when the logarithmic sheet label becomes unbounded. We first
determine which saddles can be accumulation points of a flow starting from $\mathcal C_{\rm reg}$. We then show that a trajectory with bounded $\mathcal R$ cannot avoid this conclusion by wandering through noncompact regions: it must still have a saddle as an accumulation point. The resulting bound on the saddle values will exclude both apparent possibilities.

The critical equations imply that all saddle lifts project to
\begin{equation}
    \zeta=\widetilde\zeta=2.
\end{equation}
Although the two logarithms separately carry branch labels
$(m,n)\in\mathbb Z^2$, the branches relevant here are those of their combination $\mathcal A_0$. At the base point,
\begin{equation}
    \!\!\!\Delta\mathcal A_0
    \!=\!
    2\pi i(ms_0+nt_0)
    =
    2\pi i s_0\sigma,
    ~
    \sigma=m-\frac{n}{2}
    \in\frac{1}{2}\mathbb Z.
\end{equation}
Pairs $(m,n)$ with the same $\sigma$ define the same branch of
$\mathcal A_0$, so its logarithmic sheets may be labelled directly by $\sigma$. The corresponding saddle values are
\begin{equation}
    \mathcal A_0(p_\sigma)
    =
    s_0\left(\log4+2\pi i\sigma\right)\,.
\end{equation}

We next determine which of these saddles could be approached from
$\mathcal C_{\rm reg}$. On its undeformed part,
\begin{equation}
    \operatorname{Im}\mathcal A_0
    =
    2\mathcal R.
\end{equation}
If a trajectory starting at $X_0$ on this part accumulated at
$p_\sigma$, conservation of the imaginary part and monotonicity of
$\mathcal R$ would require
\begin{equation}
    \frac{1}{2}
    \operatorname{Im}\mathcal A_0(p_\sigma)
    =
    \mathcal R(X_0)
    \leq
    \operatorname{Re}\mathcal A_0(p_\sigma).
\end{equation}
Substituting the saddle values gives $\sigma\leq0$, and therefore
\begin{equation}
    \operatorname{Re}\mathcal A_0(p_\sigma)
    \geq
    \kappa\log2.
\end{equation}
On the regulated part, one has $\operatorname{Im}\mathcal A_0<0$. A saddle with the same conserved imaginary part must then have $\sigma\leq-1/2$, and hence satisfies the even stronger bound
\begin{equation}
    \operatorname{Re}\mathcal A_0(p_\sigma)
    \geq
    \kappa(\log2+\pi)
    >
    \kappa\log2.
\end{equation}
Thus, every saddle that can be approached by a trajectory starting on
$\mathcal C_{\rm reg}$ has
\begin{equation}
    \operatorname{Re}\mathcal A_0(p_\sigma)
    \geq
    \kappa\log2.
\end{equation}

Choose $c>0$ such that
\begin{equation}
    0<3c<\kappa\log2.
\end{equation}
Suppose that a trajectory starting at $X_0\in\mathcal C_{\rm reg}$ with $\mathcal R(X_0)<2c$ never reaches $\mathcal R=2c$. Since $\mathcal R$ is monotone and bounded, it has a
finite limit
\begin{equation}
    \mathcal R(\Phi_\tau(X_0))
    \longrightarrow
    \mathcal R_\infty
    \leq
    2c.
\end{equation}
The imaginary part of $\mathcal A_0$ is conserved, so the full value of $\mathcal A_0$ remains in the compact line segment
\begin{equation}
    \mathcal A_0(\Phi_\tau(X_0))
    \in
    [\mathcal R(X_0),\mathcal R_\infty]
    +
    i\,\operatorname{Im}\mathcal A_0(X_0).
\end{equation}
Moreover,
\begin{equation}
    \!\!\int_0^\infty\!\!\!\!\!
    \frac{
        \|\nabla_g\mathcal R(\Phi_\tau(X_0))\|_g^2
    }{
        1+\|\nabla_g\mathcal R(\Phi_\tau(X_0))\|_g^2
    }
    \,\d\tau
    =
    \mathcal R_\infty\!\!-\!\mathcal R(X_0)
    <
    \infty.
\end{equation}
There must therefore exist a sequence $\tau_j\to\infty$ for which
\begin{equation}
    \|\nabla_g\mathcal R(\Phi_{\tau_j}(X_0))\|_g
    \to
    0.
\end{equation}
Set $X_j=\Phi_{\tau_j}(X_0)$. If $X_j$ has a convergent subsequence with limit $X_*$, continuity gives
\begin{equation}
    \|\nabla_g\mathcal R(X_*)\|_g
    =
    \lim_{j\to\infty}
    \|\nabla_g\mathcal R(X_j)\|_g
    =
    0.
\end{equation}
The trajectory would therefore have a lifted saddle as an accumulation point. It remains only to show that such a convergent subsequence exists, or equivalently that $X_j$ cannot leave a compact subset of the logarithmic covering space.

We first exclude the endpoint regions. Near $\zeta=0$, $\zeta=1$, and $\zeta=\infty$, the derivatives of $\mathcal A_0$ with respect to the corresponding local variables $\log\zeta$, $\log(1-\zeta)$, and $\log(1/\zeta)$ approach $s_0$, $t_0$, and $u_0$, respectively. The same statement holds for $\widetilde\zeta$. The metric chosen above is uniformly comparable to the flat metric in each of these logarithmic coordinates near the corresponding endpoint. Since $s_0$, $t_0$, and $u_0$ are all nonzero, the derivative limits above imply that $\|\nabla_g\mathcal R\|_g$ has a positive lower bound whenever either $\zeta$ or $\widetilde\zeta$ approaches $0$, $1$, or $\infty$. If several such limits are approached simultaneously, they give independent components of the gradient, so the same conclusion holds. Since $\|\nabla_g\mathcal R(X_j)\|_g\to0$, the sequence must remain away from all these endpoint regions.

Its projection therefore lies in a compact interior region. On the
logarithmic covering space, boundedness of $\mathcal A_0(X_j)$ also restricts the sequence to finitely many sheets: on this compact projected region, the value of $\mathcal A_0$ on a
fixed reference branch is uniformly bounded as the projected point
varies, while the remaining branches differ by
\begin{equation}
    2\pi i s_0\sigma,
    \qquad
    \sigma\in\frac{1}{2}\mathbb Z.
\end{equation}
Since $s_0\neq0$, only finitely many values of $\sigma$ are compatible with $\mathcal A_0(X_j)$ remaining in a compact set. Thus $X_j$ lies in a compact subset of the logarithmic covering space and has a convergent subsequence. Its limit is a lifted saddle $p_\sigma$. Conservation of the imaginary part gives
\begin{equation}
    \mathcal A_0(p_\sigma)
    =
    \mathcal R_\infty
    +
    i\,\operatorname{Im}\mathcal A_0(X_0).
\end{equation}
But every saddle that can be an accumulation point of a trajectory
starting on $\mathcal C_{\rm reg}$ satisfies $\operatorname{Re}\mathcal A_0(p_\sigma)\geq\kappa\log2$, whereas
\begin{equation}
    \kappa\log2
    \leq
    \operatorname{Re}\mathcal A_0(p_\sigma)
    =
    \mathcal R_\infty
    \leq
    2c
    <
    \kappa\log2,
\end{equation}
which is impossible. Hence every point with $\mathcal R<2c$ reaches the level $\mathcal R=2c$ after a finite flow time.

Finally,
\begin{equation}
    K
    =
    \mathcal C_{\rm reg}
    \cap
    \{\mathcal R\leq2c\}
\end{equation}
is compact because $\mathcal R\to+\infty$ along every noncompact end of $\mathcal C_{\rm reg}$. Every trajectory starting in $K$ actually crosses to $\mathcal R>2c$, since an obstruction at
$\mathcal R=2c$ would be an accessible saddle with real value
$2c<\kappa\log2$. Continuous dependence on the initial point therefore gives a neighbourhood of each point of $K$ whose trajectories cross this level within a common finite time. The compactness of $K$, together with the monotonicity of $\mathcal R$, then gives a single finite time $T$ valid for all points of $K$.

To leave the distant endpoint regions unchanged, we now multiply the flow vector field by a smooth function of $\mathcal R$ that equals one for $\mathcal R\leq2c$ and zero for $\mathcal R\geq3c$, and continue to denote the resulting flow by $\Phi_\tau$. This does not affect the argument above, since every trajectory considered there remains below $2c$ until it reaches the desired level. Moreover, $\mathcal C_{\rm reg}\cap\{\mathcal R\leq3c\}$ is compact, so only a compact part of the integration surface is moved. We have therefore constructed
\begin{equation}
    \mathcal C_+
    =
    \Phi_T(\mathcal C_{\rm reg}),
    \qquad
    \operatorname{Re}\mathcal A_0
    \big|_{\mathcal C_+}
    \geq
    2c
    >
    0.
\end{equation}

Only a compact part of the integration surface has been moved, while all its asymptotic regions are left unchanged. On the chosen
logarithmic cover of the complexified $(\zeta,\widetilde\zeta)$ space, the fixed-$y$ complex Beta integrand, with the replacements in Eq.~\eqref{eq:complexify_zeta}, is holomorphic
throughout the deformation. We now justify explicitly that the
finite-time flow cannot reach any of the loci $\zeta,\widetilde\zeta=0,1,\infty$.

The metric chosen above is complete. Indeed, for any curve $\gamma$
along which $\zeta$ approaches either $0$ or $\infty$,
\begin{equation}
    \operatorname{Length}_g(\gamma)
    \geq
    \int_\gamma
    \frac{|\d\zeta|}{|\zeta|}
    \geq
    \left|
        \Delta\log|\zeta|
    \right|
    \to
    \infty,
\end{equation}
whereas for a curve approaching $\zeta=1$,
\begin{equation}
    \operatorname{Length}_g(\gamma)
    \geq
    \int_\gamma
    \frac{|\d\zeta|}{|1-\zeta|}
    \geq
    \left|
        \Delta\log|1-\zeta|
    \right|
    \to
    \infty.
\end{equation}
The same estimates hold for $\widetilde\zeta$. Thus every curve
leaving through one of these endpoint regions has infinite
$g$-length. The pullback of this metric to the logarithmic covering space is complete as well.

The normalized flow has uniformly bounded $g$-speed, and the cutoff introduced above can only decrease it:
\begin{equation}
    \left\|
        \frac{\d X}{\d\tau}
    \right\|_g
    \leq
    \frac{
        \|\nabla_g\mathcal R\|_g
    }{
        1+\|\nabla_g\mathcal R\|_g^2
    }
    \leq
    \frac{1}{2}.
\end{equation}
Let $d_g$ denote the distance induced by $g$. If a trajectory had a finite maximal flow time $T_{\max}$, then for
$0\leq\tau_1<\tau_2<T_{\max}$,
\begin{equation}
    \begin{aligned}
        d_g\!\left(X(\tau_1),X(\tau_2)\right)
        &\leq
        \int_{\tau_1}^{\tau_2}
        \left\|
            \frac{\d X}{\d\tau}
        \right\|_g
        \d\tau
        \leq
        \frac{\tau_2-\tau_1}{2}.
    \end{aligned}
\end{equation}
It follows that $X(\tau)$ is a Cauchy curve as
$\tau\to T_{\max}$. Completeness gives a limit inside the logarithmic covering space, and the smooth flow equation then extends the trajectory beyond $T_{\max}$, contradicting maximality. The flow therefore exists for every finite time and reaches none of the loci $\zeta,\widetilde\zeta=0,1,\infty$.

The deformation is consequently supported on a compact part of the
integration surface and leaves all its asymptotic boundaries
unchanged, so it produces no additional boundary contribution. Its
fixed-$y$ integral is therefore unchanged, and $\mathcal C_+$ gives the same analytically continued complex Beta integral
$I_y(s_0,t_0)$ as $\mathcal C_{\rm reg}$. (Such a construction should not be expected at $(s,t)$ values for which the $y$ representation does not converge. Indeed, if there existed a $y$-independent admissible contour satisfying the same
endpoint bounds and $\operatorname{Re}\mathcal A_{s,t}\geq c>0$, absolute convergence would allow the two integrations to be interchanged and would produce a convergent $y$ representation, contradicting its known large-$y$ behaviour. Consistently, away from the base-point region the relevant saddle values need not have positive real part and may obstruct the flow construction.)

Keeping the same lift of the contour \footnote{For generic $(s,t)$, the branches of the two logarithms entering $\mathcal A_{s,t}$ must be followed separately, so the contour is lifted to the full logarithmic cover labelled by $(m,n)\in\mathbb Z^2$, rather than only by the effective label $\sigma$ used at the base point. The branches defining the starting complex-Beta germ select a lift of $\mathcal C_{\rm reg}$ to this cover. Once this initial lift is fixed, the homotopy-lifting property gives a unique lift of the entire contour deformation, and this lifted contour is kept fixed as $(s,t)$ varies.}, the deformation above remains valid in a sufficiently small neighbourhood of the base point. Since $\operatorname{Re}\mathcal A_0\geq2c$ on $\mathcal C_+$ and the endpoint inequalities on its unchanged noncompact tails are strict, sufficiently small variations of $(s,t)$ cannot make $\operatorname{Re}\mathcal A_{s,t}$ nonpositive anywhere on this contour. The neighbourhood can therefore be chosen
so that
\begin{equation}
    \operatorname{Re}\mathcal A_{s,t}
    \big|_{\mathcal C_+}
    \geq
    c
    >
    0.
\end{equation}
The $y$ and worldsheet integrations may consequently be interchanged in this
neighbourhood. Thus, for $(s,t)$ in a sufficiently small neighbourhood of
$(s_0,t_0)$, one obtains
\begin{align}
    \!\!\int_{\mathcal{C}_+}
    \!\Omega\,
    G_\phi(\mathcal{A}_{s,t}(\zeta,\widetilde{\zeta})) 
    =
    -\frac{1}{\pi}
    \int_{\alpha'}^{\infty}\d y\,
    \frac{\widetilde\phi_B(y)}{y^2}
    I_y(s,t).
\end{align}
This shows that the two representations define the same holomorphic germ at
the base point.

However, to connect this representation back to the analytic continuation
constructed above, we must verify that the flow deformation preserves the
original $G_\phi$ integral. Namely, after possibly shrinking the neighbourhood
$\mathcal U$ of the base point, we need
\begin{equation}
\begin{aligned}
\int_{\mathcal{C}_+}
\!\Omega\,G_\phi(\mathcal A_{s,t}(\zeta,\widetilde{\zeta}))
=
\int_{\mathcal{C}_{\rm reg}}
\!\Omega\,G_\phi(\mathcal A_{s,t}(\zeta,\widetilde{\zeta})) .
\end{aligned}
\end{equation}
This follows from the absence of pinches during the fixed lifted homotopy from $\mathcal C_{\rm +}$ to $\mathcal C_{\rm reg}$. (Starting from $\mathcal C_+$, where $\Re \mathcal A_{s,t}>0$ fixes $G_\phi$ on its first Riemann sheet, we follow $G_\phi$ continuously on its Riemann surface along the lifted homotopy to $\mathcal C_{\rm reg}$. Any simple crossing of $\mathcal A_{s,t}=0$ is understood through an infinitesimal detour that continues $G_\phi$ onto the appropriate sheet; since this branch point is soft, the detour gives no contribution in the zero-radius limit, and only a pinch with $\mathcal A_{s,t}=\mathrm d\mathcal A_{s,t}=0$ can obstruct the deformation.) Indeed, only the compact part $\mathcal C_{\rm reg}\cap\{\operatorname{Re}\mathcal A_0\leq3c\}$ is moved, and the finite-time flow sweeps out a compact region, denoted by $\mathcal W$. At the base point, a pinch of the branch locus during the deformation would require
\begin{equation}
    \mathcal A_0=0,
    \qquad
    \d\mathcal A_0=0 .
\end{equation}
The second condition is precisely the saddle condition of the flow, since for any holomorphic $\mathcal A_{s,t}$ the Hermitian metric $g$ gives $\mathrm d\mathcal A_{s,t}=0$ if and only if $\nabla_g\Re\mathcal A_{s,t}=0$. The saddle analysis above, which was used to construct the flow, also implies that every accessible saddle satisfies $\operatorname{Re}\mathcal A_0>0$, and hence cannot lie on $\mathcal A_0=0$. Therefore, 
\begin{equation}
    \min_{X\in\mathcal W}
    \left(
        |\mathcal A_0(X)|
        +
        \|\nabla_g\Re \mathcal A_0(X)\|_g
    \right)
    >
    0 .
\end{equation}
By continuity and compactness of $\mathcal{W}$, the neighbourhood $\mathcal U$ can be chosen so that, on the same lifted homotopy
\begin{equation}
    \min_{X\in\mathcal W}
    \left(
        |\mathcal A_{s,t}(X)|
        +
        \|\nabla_g\Re \mathcal A_{s,t}(X)\|_g
    \right)
    >
    0
\end{equation}
for all $(s,t)\in \mathcal U$. Thus no pinch occurs during the flow deformation throughout $\mathcal U$. Together with the analytic-continuation argument above, this completes the proof.

The same argument applies to $\mathcal{I}(s,u)$ and $\mathcal{I}(u,t)$. Therefore the Virasoro-Shapiro-like block is analytic in the Mandelstam domain
\begin{equation}
    s,t,u\notin[0,\infty),
\end{equation}
and its only first-sheet singular loci are the three physical
channel cuts, with massless poles at their endpoints.

\end{document}